\documentclass[fleqn,usenatbib]{mnras}

\usepackage{newtxtext,newtxmath}

\usepackage[T1]{fontenc}

\DeclareRobustCommand{\VAN}[3]{#2}
\let\VANthebibliography\thebibliography
\def\thebibliography{\DeclareRobustCommand{\VAN}[3]{##3}\VANthebibliography}

\usepackage{graphicx}	
\usepackage{amsmath}	
\usepackage{siunitx}    
\usepackage{hyperref}   
\usepackage{multirow, makecell}   
\usepackage[table]{xcolor}     
\usepackage{color, soul}

\title[Dynamical dark energy via flexknots]{Nonparametric reconstructions of dynamical dark energy via flexknots}

\author[A.N.~Ormondroyd et al.]{
    A.N.~Ormondroyd,$^{1,2}$\thanks{E-mail: ano23@cam.ac.uk}
    W.J.~Handley,$^{1,2}$
    M.P.~Hobson$^{1}$
    and A.N.~Lasenby$^{1,2}$
    \\
    $^{1}$Astrophysics Group, Cavendish Laboratory, J.J.~Thomson Avenue, Cambridge, CB3 0HE, UK\\
    $^{2}$Kavli Institute for Cosmology, Madingley Road, Cambridge, CB3 0HA, UK\\
    }

    \date{Accepted XXX. Received YYY; in original form ZZZ}

    \pubyear{2025}

\begin{document}
    \label{firstpage}
    \pagerange{\pageref{firstpage}--\pageref{lastpage}}
    \maketitle

    \begin{abstract}
        Recent cosmological surveys have provided unprecedented datasets that can be used to reconstruct the history of the dark energy equation of state.
        In this work, a free-form ``flexknot'' parameterisation is employed to represent $w(a)$ as a linear spline between free-moving nodes, the number of which may vary.
        By combining DESI Baryon Acoustic Oscillation measurements with Pantheon+ or DES5Y supernovae, the functional posteriors of $w(a)$ reveal an unexpected W-shaped structure.
        While the Bayesian evidence may still favour $\Lambda$CDM, the robustness of these results suggests the structure is indeed present in the data.
        The tension $R$-statistic and \textit{suspiciousness} have been marginalised over models, and demonstrate that while the reconstructions from DESI and Pantheon+ agree, DESI and DES5Y do not.
        We conclude that, while there is no smoking gun for dynamical dark energy, the structure unearthed in this work is generally too complex to be captured by the restrictive $w$CDM or CPL parameterisations.
    \end{abstract}

    \begin{keywords}
        methods: statistical -- cosmology: dark energy, cosmological parameters
    \end{keywords}



    \section{Introduction}

    The discovery of the accelerated expansion of the Universe has been regarded as one of the most profound developments in cosmology over the past quarter-century.
    Initial direct observational evidence came from measurements of distance moduli of Type Ia supernovae \citep{SupernovaSearchTeam:1998, SupernovaCosmologyProject:1998}, and confirmation was subsequently obtained from other independent probes such as the Cosmic Microwave Background (CMB) anisotropies \citep{planck18vi}, Baryon Acoustic Oscillations (BAO) \citep{sdss, yes}, and various other probes \citep{act, sixpercent, yes, lss, stellarage}.
    Despite forming the backbone of the standard model of cosmology, the cosmological constant is confronted with several theoretical challenges, notably the fine-tuning and coincidence problems \citep{Weinberg:1988cp, Zlatev:1998tr}.

    In the standard picture, dark energy is taken to be a constant energy density with an equation of state parameter $w = p/\rho = -1$, which was used by Einstein as a way to achieve a  static universe, albeit an unstable one \citep{einstein1917, einstein1917centenary}.
    Since the 1990s, dark energy has been used to account for the observations of late-time accelerated cosmological expansion.
    Despite this success, little progress has been made to determine the particular nature of dark energy beyond its large-scale effects.

    These developments have motivated continued scrutiny of the dark energy sector, as subtler signatures are now revealed by new datasets and may elude traditional analyses.
    In particular, the possibility that the dark energy equation of state may be a constant other than $-1$, or even evolve with time, is being explored as a potential avenue for new physics.
    The two most popular phenomenological parameterisations are one in which $w$ is allowed to vary as a single constant ($w$CDM), and the Chevallier-Polarski-Linder (CPL) form \citep{Chevallier:2000qy, Linder:2002et}, in which $w$ evolves linearly with the scale factor according to $w(a) = w_0 + (1-a) w_a$.

    When $w$CDM is employed, the DESI BAO papers have reported that the posterior is in agreement with $\Lambda$CDM (i.e. a value consistent with $-1$;  \cite{desiiii, desiiv, desivi}).
    However, when the Chevallier-Polarski-Linder (CPL) equation of state is applied, DESI combined with Pantheon+ \citep{pantheonplus} or particularly DES5Y \citep{des5y} supernovae have been seen to deviate from $\Lambda$CDM (at least without the ``correction'' applied by \cite{georgedes5y}).
    These contrasting outcomes suggest that conventional parametric forms may be too restrictive to capture fully any evolution of dark energy; see \cite{dedataoverview} for an overview of how different datasets constrain CPL dark energy.

    The discrepancy raises an important question: could traditional parametric forms be too restrictive?
    To explore this possibility, a model-agnostic approach is adopted.
    Rather than presupposing a particular functional form for $w$, an alternative non-parametric route using flexknots is followed.
    Flexknots are a flexible parameterisation of one-dimensional functions without smoothness restrictions, thereby allowing arbitrarily sharp features to be reconstructed.
    This approach permits sudden changes in $w$ to emerge.
    The goal is to reconstruct the dark energy equation of state in a manner that faithfully captures the underlying structure in the datasets, alone and combined.
    This technique has previously been used successfully to reconstruct the history of the dark energy equation of state from CMB data \citep{sonke, devazquez}, the primordial power spectrum \citep{pkhandley, pkvazquez, pkknottedsky, pkcore, pkplanck13, pkplanck15}, the cosmic reionisation history \citep{flexknotreionization, heimersheimfrb}, galaxy cluster profiles \citep{flexknotclusters}, and the $\SI{21}{\centi\metre}$ signal \citep{heimersheim21cm, shen}.
    It is not being suggested that dark energy equation of state parameter evolves strictly as a series of straight lines; rather, these reconstructions reveal the underlying structure present in the data --- features that any viable physical model for dark energy must ultimately accommodate.

    In this study, BAO and supernovae are systematically combined and evaluated using the tension $R$-statistic and suspiciousness \citep{lemos, hergt, balancingact}, so it can be assessed whether the resulting structures are being driven by disagreement between the datasets.

    This paper is organised as follows. In Section~\ref{sec:data}, we describe the observational datasets used. In Section~\ref{sec:methods}, the methodology for reconstructing dark energy with flexknots is outlined, along with a brief introduction into tension statistics, then the sampling strategies and priors employed in this work. In Section~\ref{sec:results}, datasets are considered in turn to assess which are responsible for features found in $w(a)$. We present our conclusions in Section~\ref{sec:conclusions}, together with suggestions for future work. We also include several appendices containing mathematical details.

    \section{Data}\label{sec:data}

    Given the surprising structure in the $w(a)$ reconstructions, and as the calculations for BAO and Type Ia supernovae are relatively straightforward, it was decided to create a pared-down pipeline. 
    This is to ensure that the results are not caused by an obscure detail in the pipeline, and to allow for a more detailed investigation of the data.
    Detail is included in Appendix~\ref{apx:pipeline}.

    \subsection{DESI BAO}

    Baryon Acoustic Oscillations (BAO) are a primary cosmological probe. They arise from periodic fluctuations in the density of visible baryonic matter in the universe, echoes of sound waves which propagated through the early universe.
    These oscillations were frozen at recombination when the universe had cooled sufficiently for protons and electrons to combine into neutral hydrogen, and photons decoupled from baryons, thereby providing a standard ruler for measuring the expansion history of the universe.

    BAO datasets are compressed into the ratio of either the transverse comoving distance $D_\text M(z)$ or Hubble distance along the line of sight $D_H(z)$ to the sound horizon at decoupling $r_\text d$. In a flat universe:
    \begin{equation}
        \begin{aligned}
            \frac{D_\mathrm M(z)}{r_\mathrm d} &= \frac{c}{r_\mathrm d}\int_0^z\frac{\mathrm dz'}{H(z')} = \frac{c}{r_\mathrm dH_0}\int_0^z\frac{\mathrm dz'}{h(z')}  &(\Omega_k=0)\text, \\
            \frac {D_\mathrm H(z)}{r_\mathrm d} &= \frac{c}{r_\mathrm dH(z)} = \frac{c}{r_\mathrm dH_0}\frac{1}{h(z)} \text, & h(z) = \frac{H(z)}{H_0}\text.
        \end{aligned}
    \end{equation}

    DESI DR1 supplies BAO measurements from several galaxy samples: the Bright Galaxy Sample (BGS), the Luminous Red Galaxy Sample (LRG, split into three disjoint redshift ranges), the Emission Line Galaxy Sample (ELG), and the Quasar Sample (QSO).
    Since the redshift ranges of the third LRG and ELG overlap, optimal BAO information is obtained by combining the two samples \citep{desiiii}.
    In cases where the signal-to-noise ratio is insufficient to measure both distances (as in the BGS and QSO samples), the volume averaged $D_\mathrm V(z) = (zD_\mathrm M^2 D_\mathrm H)^{\frac 1 3}$ is reported instead.
    These data are used as they appear in Table~1 of \cite{desivi}.
    Moreover, as the two LRG data points at $z=0.510$ and $z=0.706$ are of particular interest, an investigation is undertaken into substituting these with the corresponding measurements from the Sloan Digital Sky Survey 2016 data release (SDSS, \cite{sdss}).

    \subsection{Pantheon+ and DES5Y Type Ia supernovae}

    Type Ia supernovae provide complementary constraints through measurements of their luminosity distance $D_\mathrm L(z)$, which is related to the distance modulus $\mu$:
    \begin{equation}
        \begin{aligned}
            D_\mathrm L(z) &= (1+z_\mathrm{hel})c\int_0^{z_\mathrm{HD}}\frac{\mathrm dz'}{H(z')}\text{,} & \mu(z) &= 5\log_{10}\left(\frac{D_\mathrm L(z)}{\SI{10}{pc}}\right) \\
            && &= m_B - M_B\text.
        \end{aligned}
    \end{equation}
    The observed heliocentric redshift $z_\mathrm{hel}$ is distinct from the redshift due to the Hubble flow $z_\mathrm{HD}$, which is corrected for the peculiar velocity of the distant galaxy hosting the supernova.
    $M_B$ and $m_B$ are the absolute and apparent magnitude of Type Ia supernovae. The apparent magnitudes and redshifts are reported in the datasets, and the absolute magnitude and the Hubble constant are degenerate.
    Both of these have been analytically marginalised out of the likelihood, with mathematical detail included in Appendix~\ref{apx:absolute}.
    Consequently, $\Lambda$CDM only has a single sampled parameter: $\Omega_\mathrm m$.
    Cepheid variable distance calibration is not used in this work.

    We use the Pantheon+ and DES5Y supernovae data sets \citep{pantheonplus, des5y}.
    With Pantheon+ supernovae, we consider lower redshift cut-offs of $z_\mathrm{HD}=0.01$, as fixed in the \texttt{Cobaya} source code\footnote{\href{https://github.com/CobayaSampler/cobaya/blob/5c605d39ab7e0a13f39312b123ffd5fc3d6a2aba/cobaya/likelihoods/sn/pantheonplus.py\#L38}{github.com/CobayaSampler/cobaya}} \citep{cobaya1, cobaya2}, and $z_\mathrm{HD}=0.023$ to match the $\mathrm{SH}_0\mathrm{ES}$ analysis and other analyses of the effect of Ia supernovae on dark energy \citep[e.g.,][]{sh0es, pantheonz23, toby}.
    It is found that neither choice qualitatively alters the headline results; a comparison is provided in Appendix~\ref{apx:iacutoff}.

    \section{Methods}\label{sec:methods}

    \subsection{Flexknot dark energy}

    \begin{figure}
        \centering
        \includegraphics[width=\columnwidth]{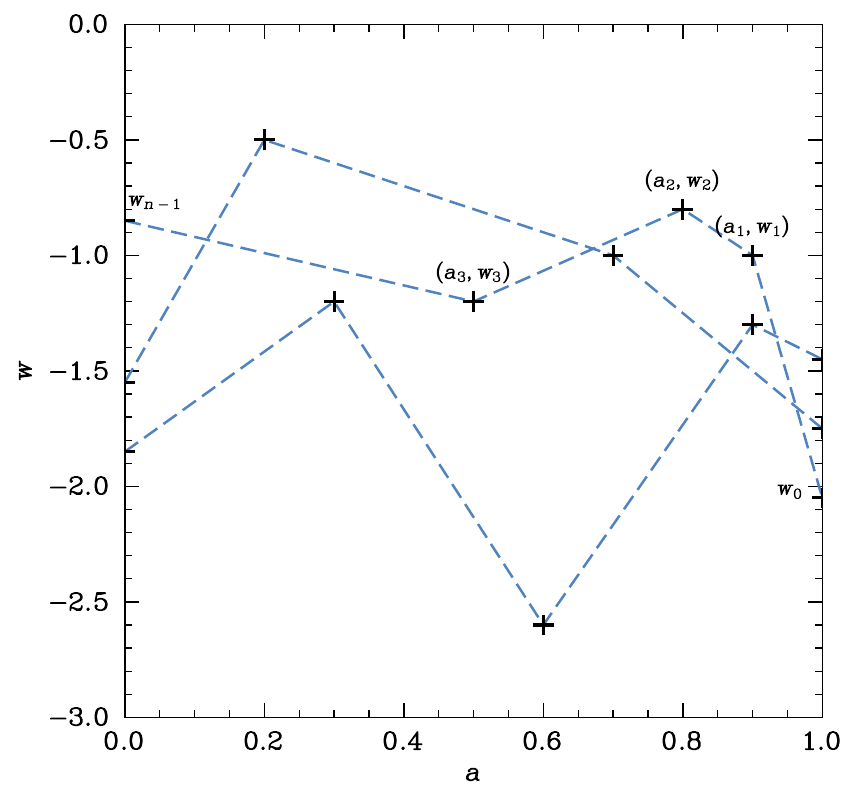}
        \caption{
            Examples of dark energy equation of state flexknots.
            The end knots are fixed at scale factors of $0$ and $1$, while the others are free to move anywhere between them, with the restriction that they remain sorted in the correct order.
            Unlike previous usage, knots are numbered from right to left in decreasing scale factor, so that $a_0$ is the present day.
            In this scheme, $w_0$ is consistent with other parameterisations.
        }
        \label{fig:egflexknot}
    \end{figure}

    \begin{figure*}
        \centering
        \includegraphics[width=0.48\textwidth]{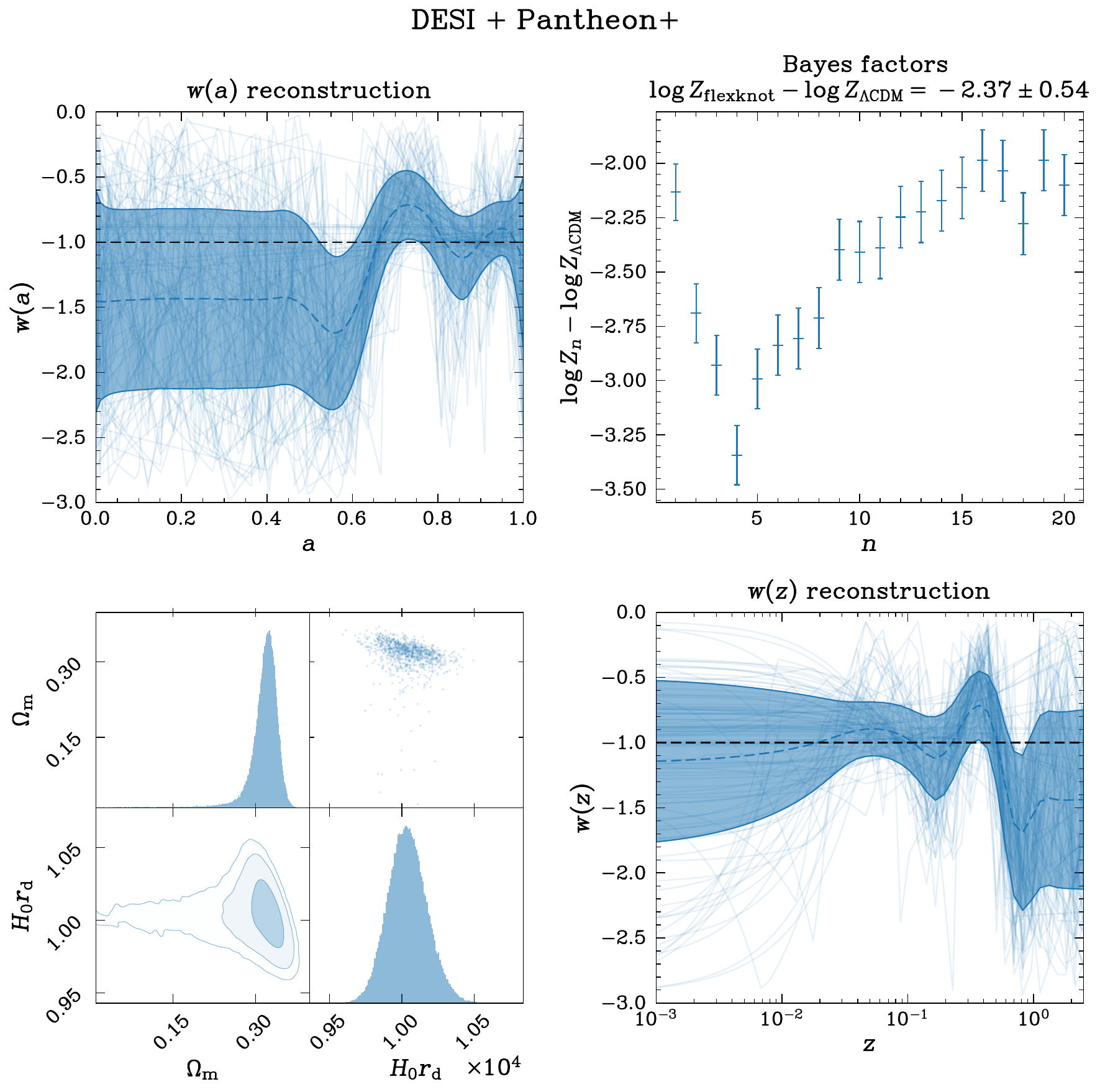}
        \includegraphics[width=0.48\textwidth]{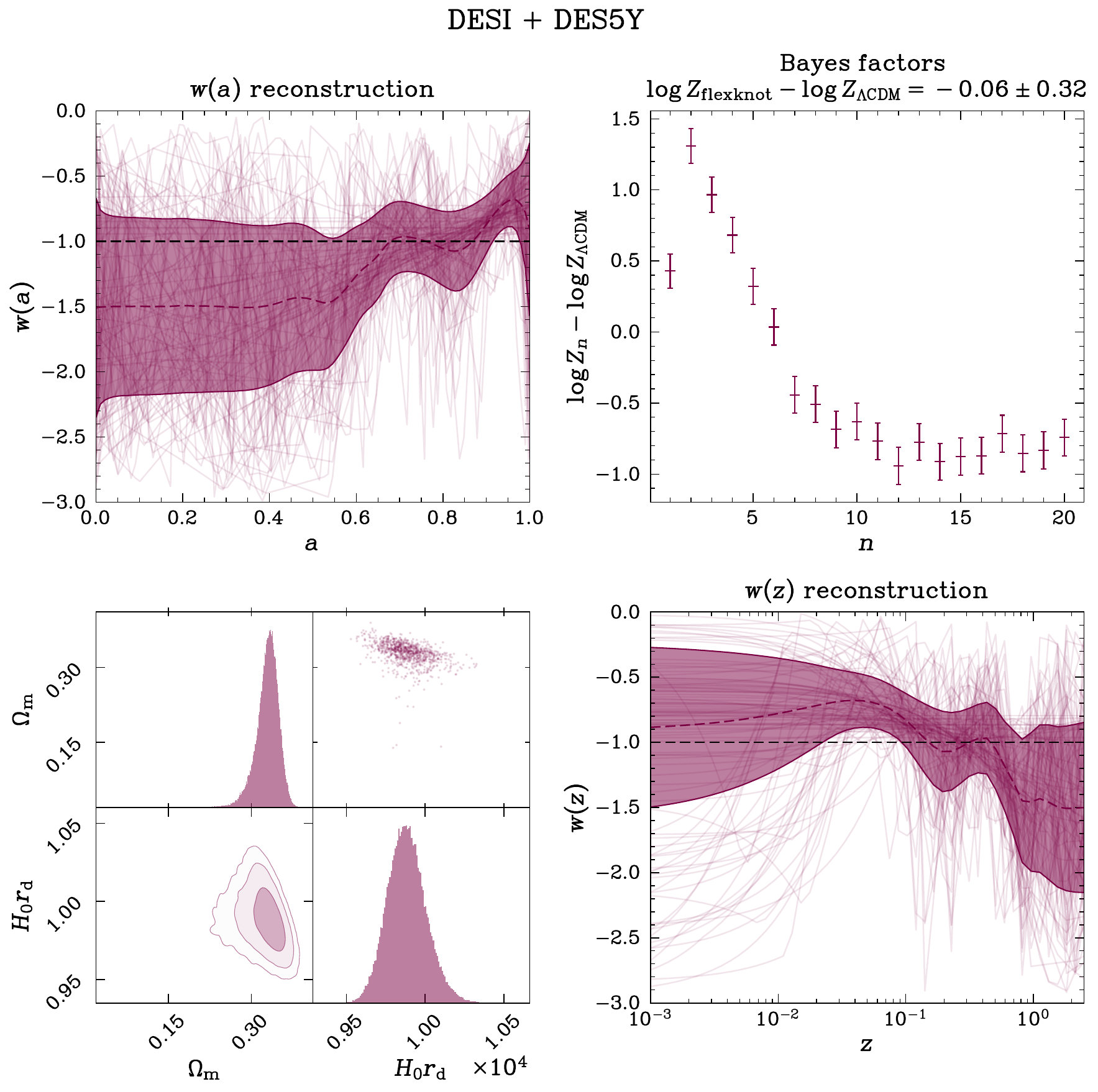}
        \caption{
            Left four panels: dark energy flexknot reconstruction using DESI and Pantheon+.
            The upper-left panel shows posterior samples of the evolution of $w(a)$.
            The blue dashed line represents their mean and the shaded region corresponds to the $1\sigma$ contour, while the black dashed line is $\Lambda$CDM.
            The upper-right panel shows the evidence for the $n^\mathrm{th}$ flexknot, normalised by the evidence of $\Lambda$CDM.
            The lower-left panel displays the posteriors for the parameters on which the likelihoods principally depend.
            The lower-right panel presents the same reconstruction as the upper-left, transformed into $w(z)$.
            No value of $n$ is favoured over $\Lambda$CDM, though there is little to choose between it and $w$CDM.
            Nonetheless, structure is evident in the flexknot $w$ posterior. 
            Right four panels: similar to left four with Pantheon+ replaced by DES5Y.
            In this case $n=2$ is the favoured model, and the evidences tail off for higher $n$.
            This is apparent in the $w(a)$ posterior, where by eye it is clear a straight line can fit the $1\sigma$ flexknot contour.
        }
        \label{fig:desipan}
    \end{figure*}

    \begin{figure*}
        \centering
        \includegraphics[width=0.48\textwidth]{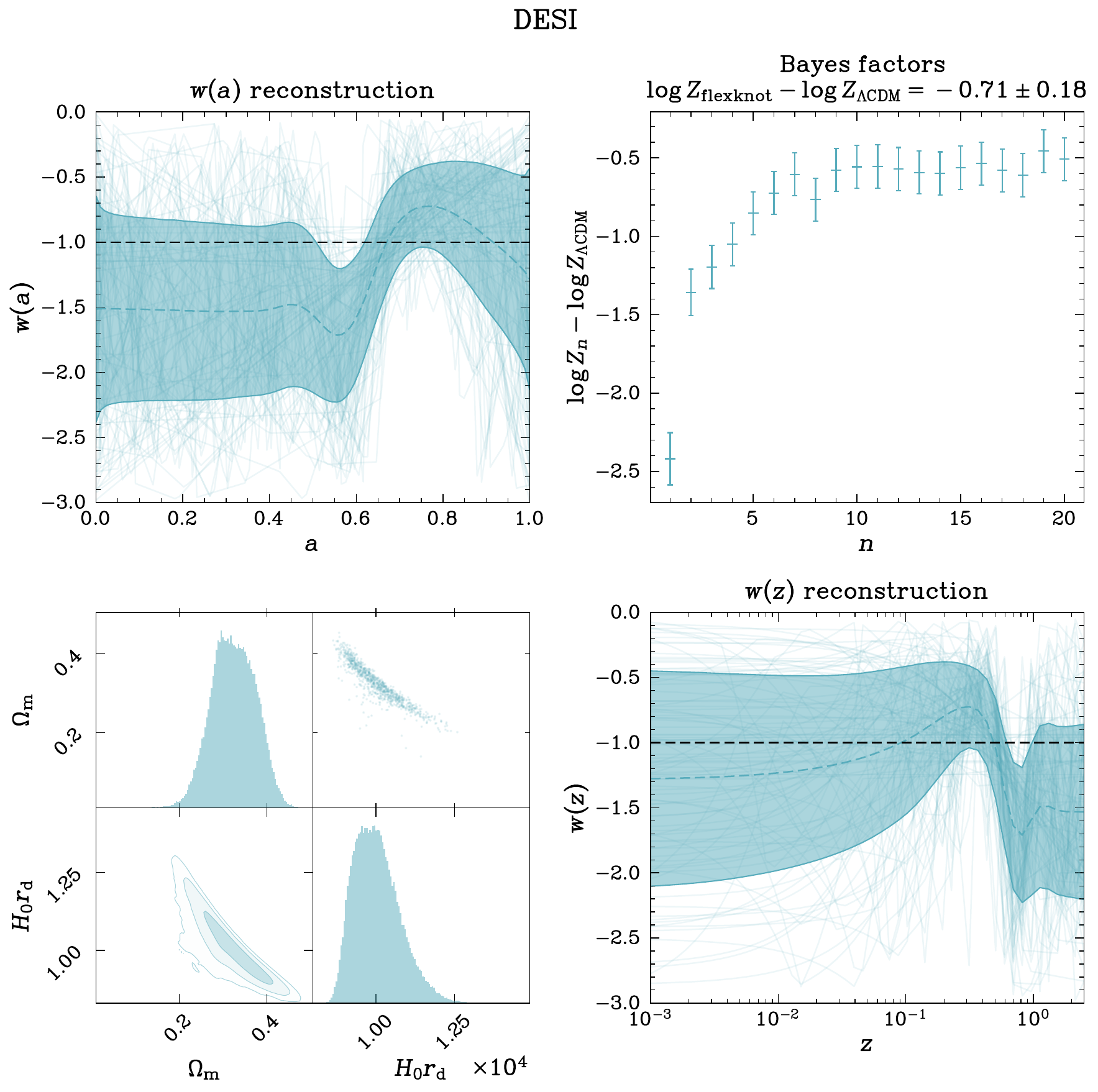}
        \includegraphics[width=0.48\textwidth]{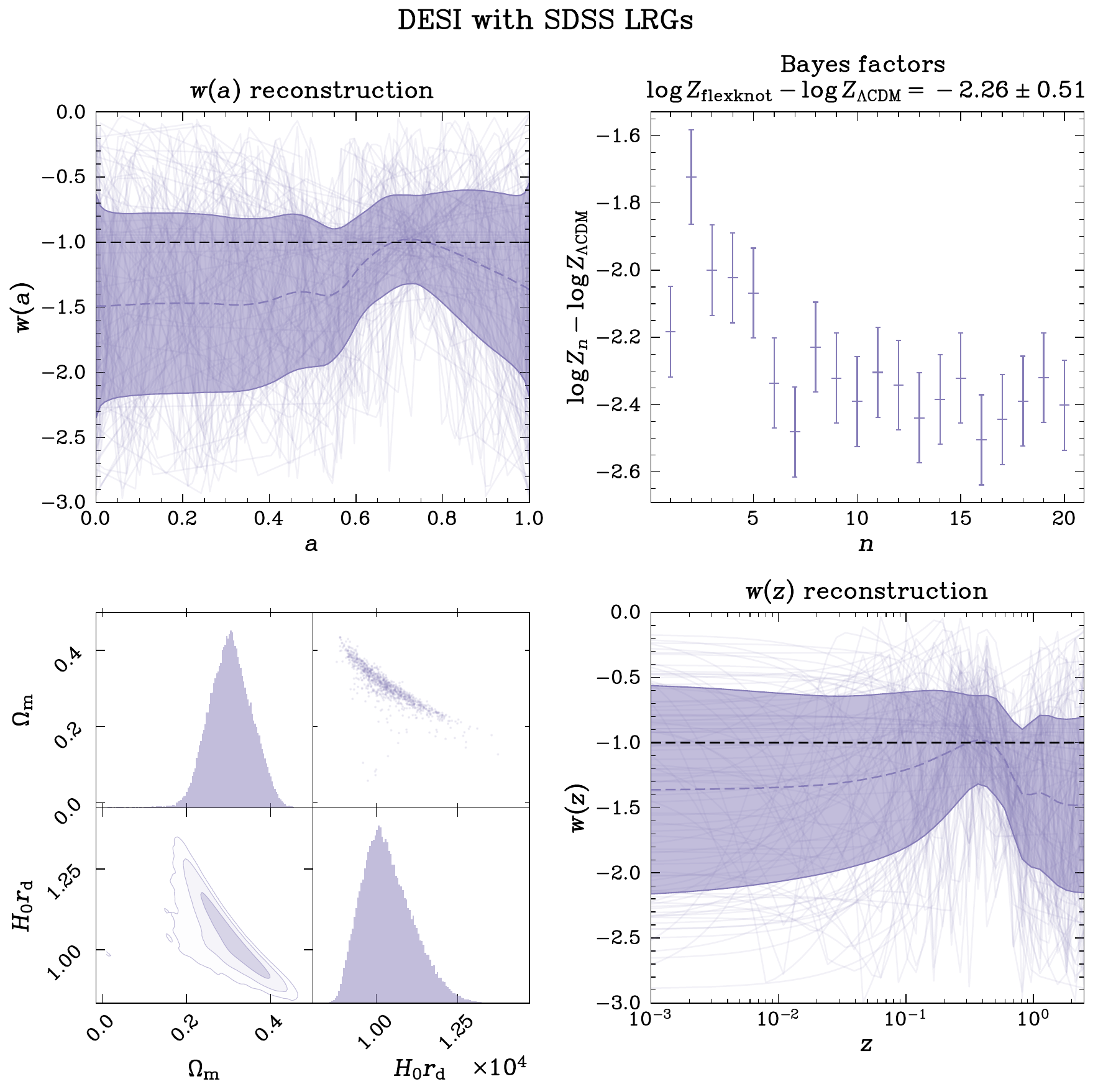}
        \caption{
            Similar to Figure~\ref{fig:desipan}, but with BAO data only.
            Left four panels: DESI is responsible for the higher redshift (low-$a$) feature.
            The evidence plateau suggests that no further structure would be found by increasing $N$ further.
            Right four panels: DESI data with the two LRG data points replaced by the equivalent SDSS DR16 data points.
            Constant $w$ is on-par with models with more knots, with $w_0=-1.03\pm0.14$.
            By contrast, the original DESI data reject $w$CDM, albeit with a similar $w_0=-1.01\pm0.14$.
            It can be seen that the DESI $1\sigma$ contour cannot admit a constant $w$, but the SDSS LRGs could. 
        }\label{fig:desisdss}
    \end{figure*}

    \begin{figure*}
        \begin{center}
            \includegraphics[width=0.95\textwidth]{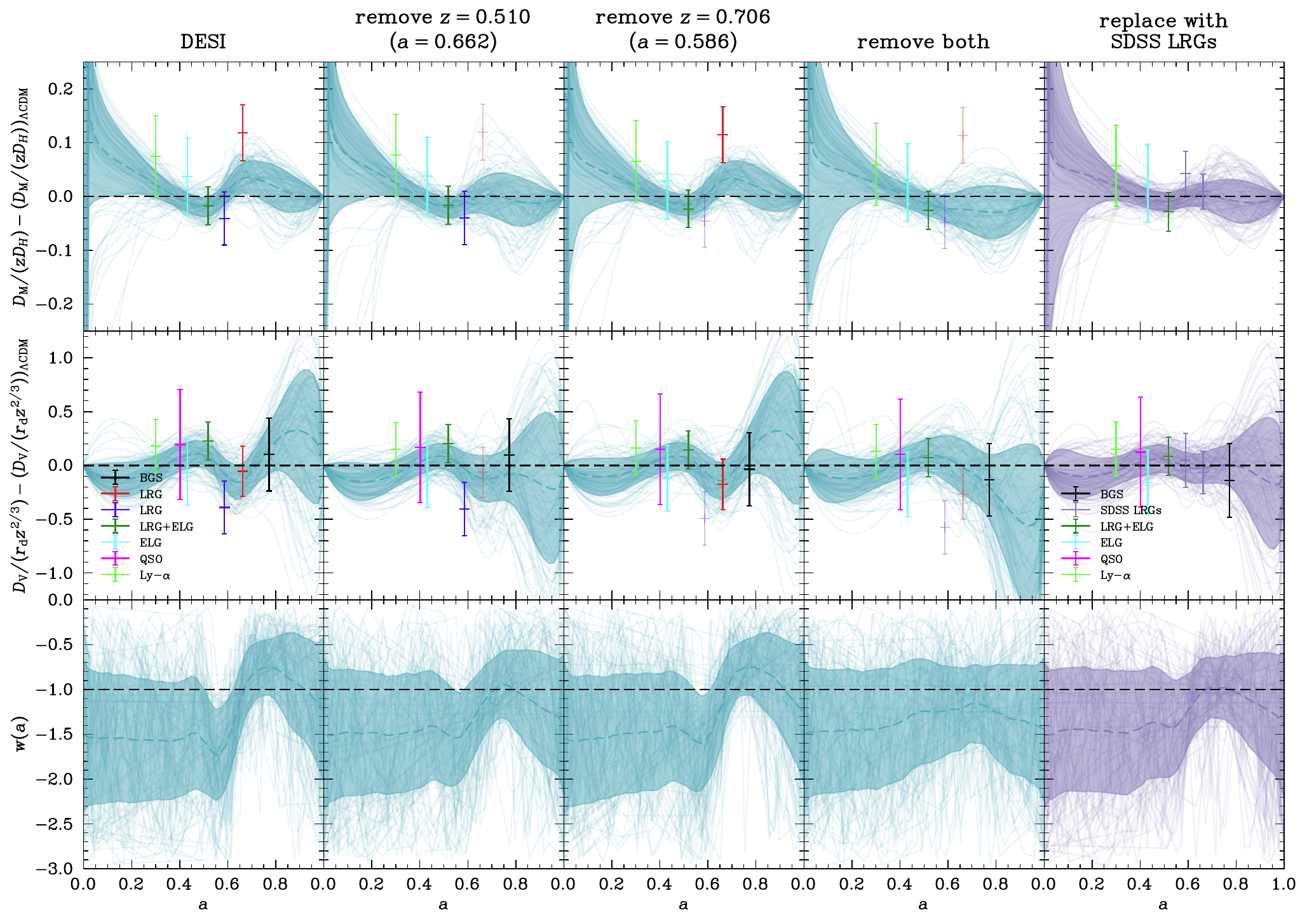}
        \end{center}
        \caption{
            Functional posterior samples of cosmological distances relative to the best-fit $\Lambda$CDM for DESI.
            The sample mean is shown as a dashed line.
            The ratios of distances correspond to Figure~1 of \protect\cite{desivi}, here plotted against scale factor for easier comparison with the $w(a)$ reconstruction.
            The left column uses the complete DESI DR1, while the following columns are produced by removing each of the LRG data points in turn; unused points are shown translucent.
            The red $z=0.510$ point is the primary driver for the period of quintessance, while removal of the dark blue $0.706$ point makes little difference to the dark energy reconstruction and its corresponding distance calculations.
            The final column replaces those LRGs with the matching values from SDSS DR16, resulting in a very similar reconstruction to that of the second column.
        }\label{fig:baodistances}
    \end{figure*}

    In order to represent a truly free-form function, the dark energy equation of state is parameterised as a linear spline defined by a series of nodes in $(a, w)$ space.
    The function is expressed as:
    \begin{equation}
        \begin{aligned}
            w(a) &= \mathrm{flexknot}(a, \theta_w)\text{,}\\
            \theta_w &= (a_{n-1}, a_{n-2}, \dots, a_1, a_0, \\
            &\qquad w_{n-1}, w_{n-2}, \dots, w_1, w_0)\text{.}
        \end{aligned}
    \end{equation}
    The final parameter is $n$, the number of knots in the flexknot.
    Throughout this paper, $n$ refers to the total number of knots \textbf{including} the endpoints.
    For example: $n=2$ corresponds to a single linear segment similar to the CPL parameterisation, and a single knot ($n=1$) corresponds to $w$CDM.
    The capital letter $N$ is used to indicate the maximum number of knots in use, and latin indices (e.g. $n=i$ or $j$) represent a specific value of $n$.

    There are several benefits of the flexknot approach the CPL parameterisation.
    Firstly, a two-knot flexknot reproduces the CPL parameterisation, though with a different prior.
    Also, the nature of the CPL form implies that, for example, an early period of phantom dark energy combined with a preference for $\Lambda$CDM at intermediate redshift would force quintessence at late times, even with no further information.
    Flexknots have no such restriction, and will return to the prior in regions of no structure.

    A similar approach to flexknots was employed recently in \cite{2025arXiv250207185B}, where $H(z)$ was reconstructed as a piecewise constant.
    Unlike that work, the knots joining the segments of the flexknot reconstruction are permitted to shift freely to the redshifts where structure lies in the data.
    Flexknots also constitute an established technique for reconstructing one-dimensional functions in cosmology.

    \subsection{Evidence and Tension}\label{sec:tension}

    Tension between uncorrelated datasets, here denoted $A$ and $B$, is quantified by considering the tension $R$-statistic.
    $R$ can be expressed either as the probability of one dataset given the other or, alternatively, as the ratio of the evidence for the null hypothesis and the alternative hypothesis:
    \begin{equation}
        \begin{aligned}
            R &= \frac{Z(A, B)}{Z(A)Z(B)} = \frac{Z(A|B)}{Z(A)} = \frac{Z(B|A)}{Z(B)} \\
            &= \frac{Z(\text{datasets }A\text{ and }B\text{ fit one universe together})}{Z(\text{datasets }A\text{ and }B\text{ fit one universe each})} \text.
        \end{aligned}
        \label{eq:ratio}
    \end{equation}
    Following the work of \cite{lemos, hergt} and \cite{balancingact}, the Kullback-Leibler (KL) divergence from prior to posterior is rearranged using Bayes' theorem:
    \begin{equation}
        \begin{aligned}
            \mathcal P(\theta | D) Z(D) &= \mathcal L(D | \theta) \pi(\theta) \implies \frac{\mathcal P}{\pi} = \frac{\mathcal L}{Z} \\
            \mathcal D_\mathrm{KL}(\mathcal P || \pi) &= \int \mathcal P(\theta) \log\frac{\mathcal P(\theta)}{\pi(\theta)} \mathrm d\theta = \int \mathcal P \log \frac{\mathcal L}{Z}\mathrm d\theta\\
            &= \langle \log \mathcal L \rangle_\mathcal P - \log Z \text,\\
            \therefore \log Z &= \langle \log \mathcal L \rangle_\mathcal P - \mathcal D_\mathrm{KL}(\mathcal P || \pi)\text.
        \end{aligned}
    \end{equation}
    The log-evidence is thereby expressed as the sum of the likelihood averaged over the posterior and the strongly prior-dependent KL divergence. Under the assumption that the posterior is not strongly affected by reasonable choices of prior, the first term is approximately prior-independent.
    Taking the natural logarithm of equation \ref{eq:ratio} and adding KL divergences to each log-evidence transforms the Bayes' ratio into the approximately-prior-independent suspiciousness statistic:
    \begin{equation}
        \log S = \langle \log \mathcal L \rangle_{\mathcal P(A, B)} - \langle \log \mathcal L \rangle_{\mathcal P(A)} - \langle \log \mathcal L \rangle_{\mathcal P(B)} \text.
        \label{eq:suspiciousness}
    \end{equation}
    
    Appendix~\ref{apx:tension} generalises these formulae to marginalise both $R$ and $\log S$ over multiple models.
    This approach is more robust than relying solely on visual assessments from posterior corner plots for evaluating dataset agreement.

    In the case of correlated datasets, the likelihood for the pair of datasets is not given by the product of the two individual likelihoods.
    The analogous process for tension between correlated datasets is demonstrated in \cite{lemoscorr, balancingact}.
    Because the supernova datasets share common objects and BAO experiments may be interdependent, tension calculations are restricted to combinations of BAO and Type Ia supernovae that can be treated as uncorrelated.

    \begin{figure*}
        \centering
        \includegraphics[width=0.48\textwidth]{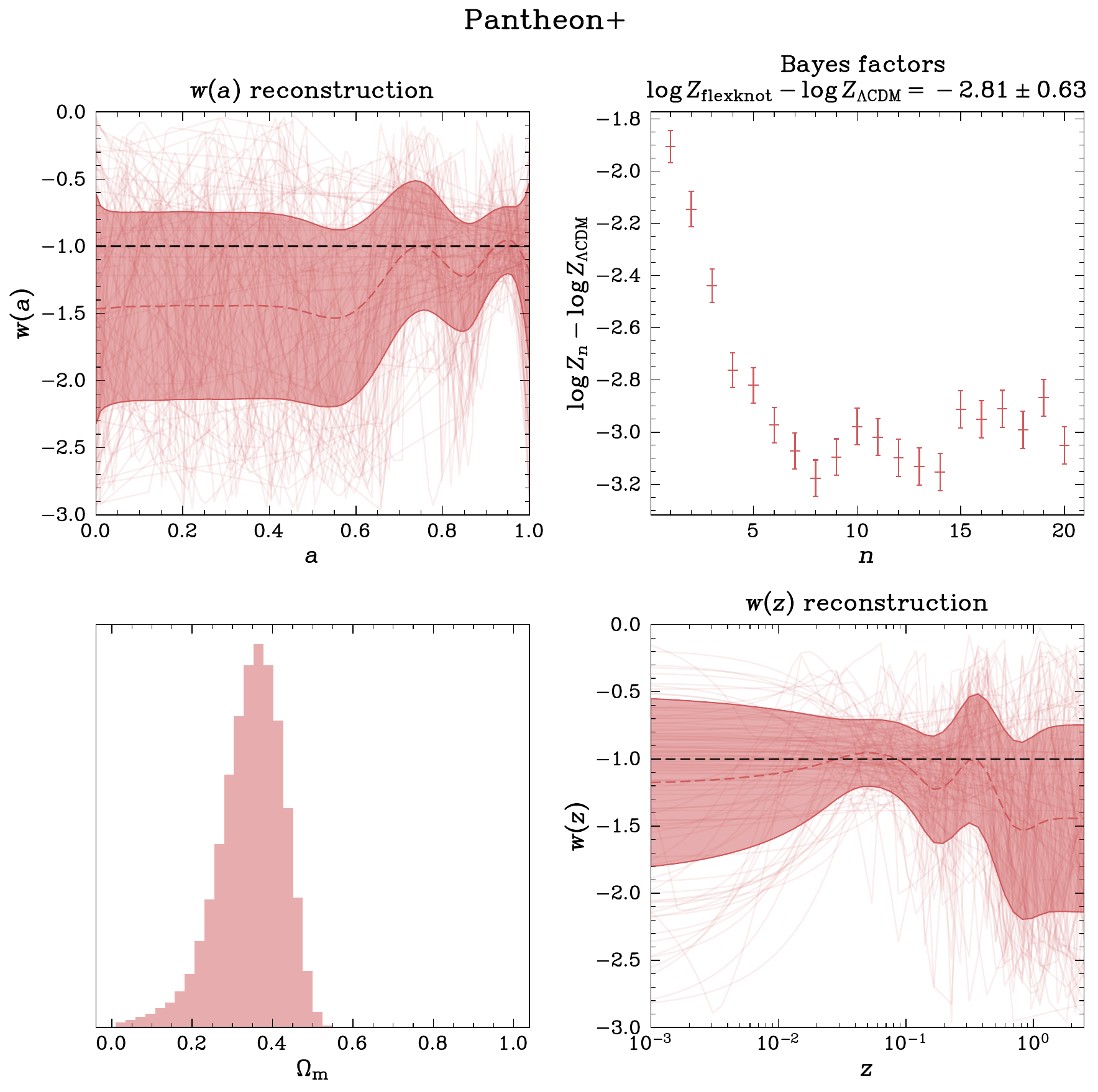}
        \includegraphics[width=0.48\textwidth]{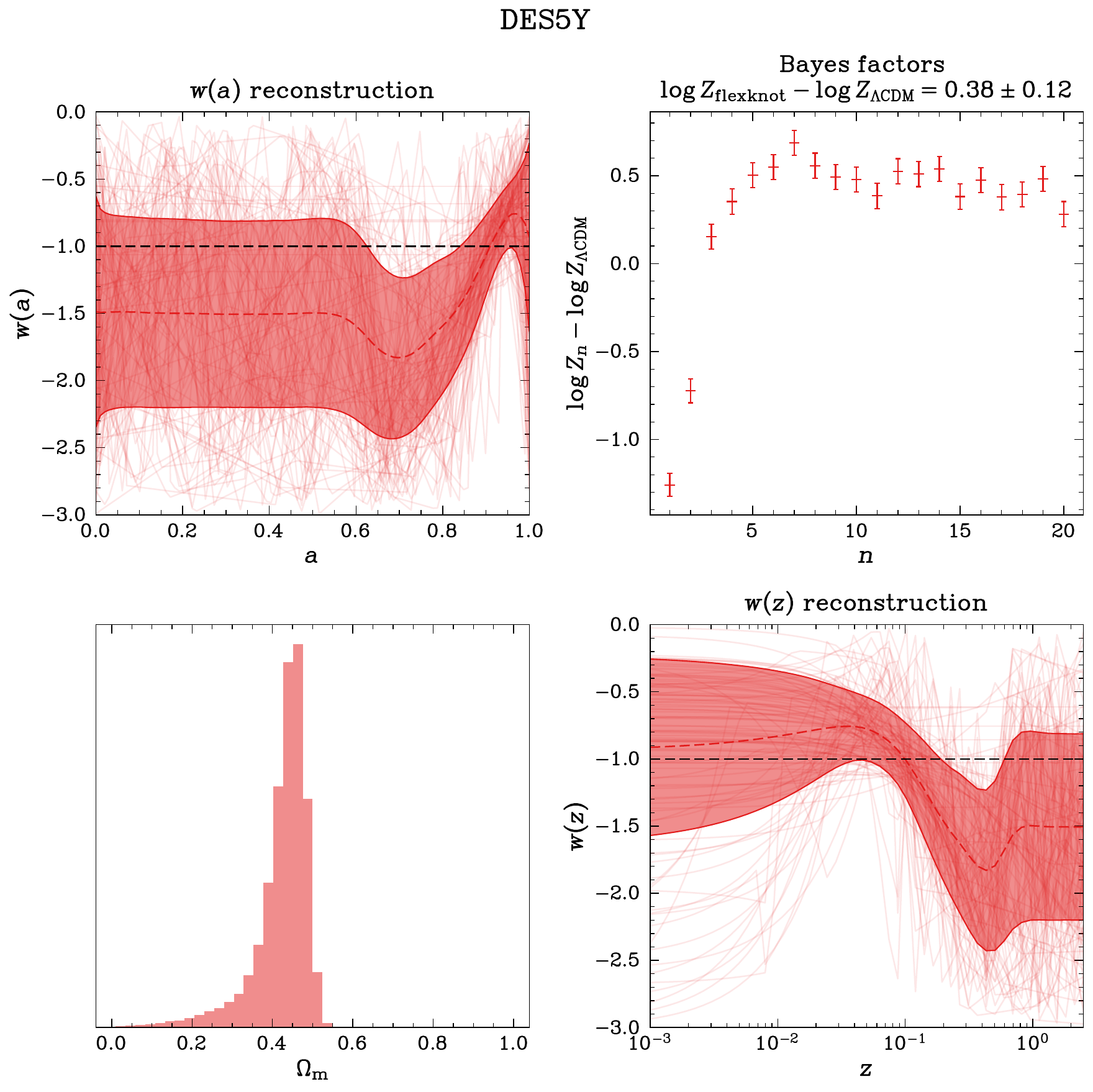}
        \caption{
            left: Similar to Figure~\ref{fig:desisdss}, but using Pantheon+ data only.
            Pantheon+ is responsible for the lower redshift (high-$a$) feature from Figure~\ref{fig:desipan}.
            Unlike DESI, Pantheon+ prefers fewer knots.
            Right: The same as left, now using DES5Y data.
            Here, the Bayesian evidence slightly favours flexknot dark energy over $\Lambda$CDM, particularly when three or more knots are employed.
        }\label{fig:pantheon}
    \end{figure*}

    \begin{figure*}
        \centering
        \includegraphics[width=0.96\textwidth]{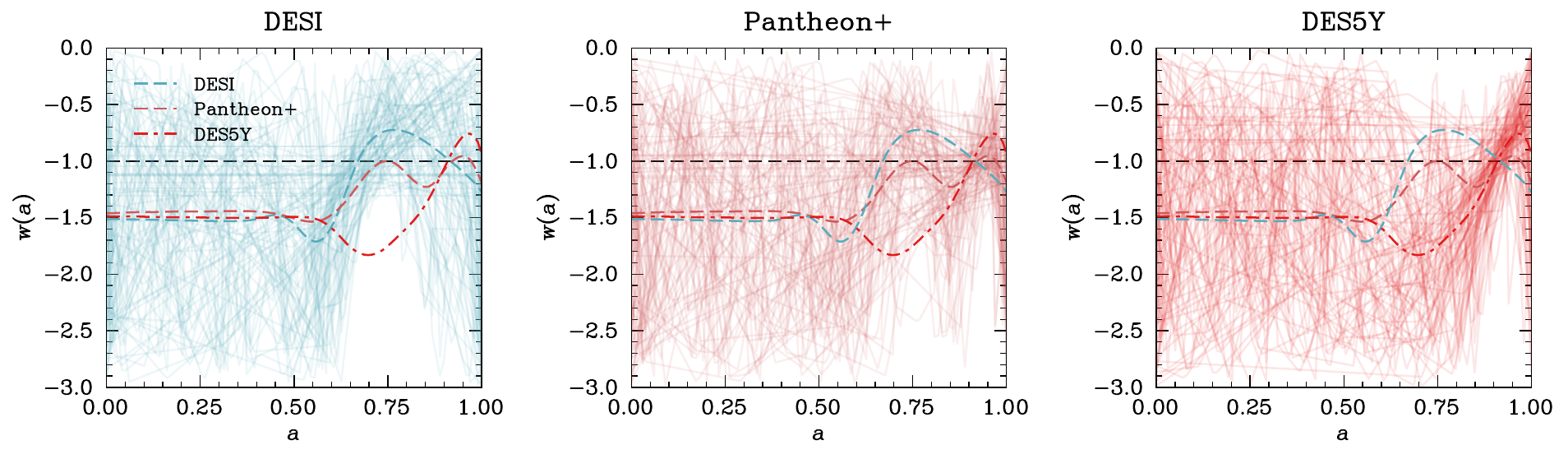}
        \caption{
            Comparison of individual reconstructions from DESI, Pantheon+ and DES5Y, using up to twenty knots ($N=20$).
            The mean of each reconstruction is displayed on all panels as dashed or dot-dashed lines.
            The $1\sigma$ contours have been omitted for clarity.
            Both the DESI and Pantheon+ reconstructions show a local maximum at $a\approx0.75$, whereas the DES5Y reconstruction exhibits a deep phantom dip.
            This discrepancy highlights the tension between DES5Y and BAO.
        }\label{fig:annaplot}
    \end{figure*}

    \subsection{Sampling and Priors}

    \begin{table}
        \centering
        \rowcolors{2}{}{gray!25}
        \begin{tabular}{|l|c|}
            \hline
            Parameter & Prior \\
            \hline
            $n$ & $[1, 20]$ \\
            $a_{n-1}$ & $0$ \\
            $a_{n-2}, \dots, a_1$ & sorted($[a_{n-1}, a_0])$ \\
            $a_0$ & $1$ \\
            $w_{n-1}, \dots, w_0$ & $[-3, -0.01]$ \\
            $\Omega_\mathrm m$ & $[0.01, 0.99]$ \\
            $H_0r_\mathrm d$ (DESI)& $[3650, 18250]$ \\
            $H_0$ (Ia) & $[20, 100]$ \\
            \hline
        \end{tabular}
        \caption{
            Cosmological priors used in this work.
            Fixed values are indicated by a single number, while uniform priors are denoted by brackets.
            As BAO only depend on the product $H_0r_\mathrm d$, and supernovae depend on $H_0$, those parameters are only included as necessary.
        }
        \label{tab:priors}
    \end{table}

    Posterior samples and evidences were obtained using the nested sampling algorithm \texttt{PolyChord}, and tension calculations were performed using \texttt{anesthetic} \citep{skilling2004, polychord1, polychord2, anesthetic}.
    A branch of \texttt{fgivenx} was used to produce the functional posterior plots \citep{fgivenx}.

    Since BAO distance ratios depend on the product $H_0r_\mathrm d$, the BAO-only sampling runs are performed by sampling this product rather than $H_0$ and $r_\mathrm d$ separately.
    Type Ia supernovae luminosity distances require only $H_0$, but since it can be analytically marginalised (see Appendix~\ref{apx:h0}), it is not sampled directly, but an explicit prior is still imposed.
    The absolute magnitude $M_B$ is not assigned an explicit prior volume; consequently, evidences are normalised with respect to $\Lambda$CDM, technically making them Bayes factors.

    Early testing involved up to thirty knots; however, it was found that using more than twenty did not significantly affect the shape of the $w(a)$ posteriors.
    For each $n$, only the necessary $(a_i, w_i)$ to describe the flexknot are sampled.
    Parameters that do not affect the likelihood leave the analytical evidence unchanged, so mathematically the extra parameters are treated as present yet inactive.
    Unused $a_i$ are omitted from the sorted prior, since their inclusion would alter the effective prior for the active parameters.
    This treatment is discussed in more detail in Appendix~\ref{apx:anthony}.

    Table~\ref{tab:priors} lists the cosmological priors used in this work. These have been chosen to be consistent with those used in \cite{desicrossingstatistics}, which provides a similar reconstruction of $w(z)$ but using Chebychev polynomials.

    \section{Results}\label{sec:results}

    Together, the BAO and supernova datasets reveal a W-shaped structure in the dark energy equation of state.
    The left four panels of Figure~\ref{fig:desipan} show this distinctive structure in $w(a)$ from the combined DESI and Pantheon+ data, which requires around ten knots or more for faithful recreation.
    $\Lambda$CDM is favoured over the flexknots, however, with a log-evidence ratio of $-2.37\pm0.54$.

    When DES5Y is used instead of Pantheon+, features occur at comparable scale factors, but these are better aligned to admit a CPL dark energy model.
    This is demonstrated in the right four panels of Figure~\ref{fig:desipan}, where the overall evidence is less strongly in favour of $\Lambda$CDM, with a log-evidence ratio of $-0.06\pm0.32$.
    This does not tell the full story, as the evidence for $n=1$, $2$, $3$, and $4$ knots are each greater than $\Lambda$CDM, before larger $n$ tail off to a constant.
    This is consistent with the functional posteriors: flexknot $w(a)$ for DESI+DES5Y visually admits a straight line, in contrast to the Pantheon+ case.

    The bottom-right panel of Figure~\ref{fig:desipan} exhibits oscillating features similar to those in Figures 1 and 2 of \cite{quintom}, which are produced using Gaussian processes.
    However, the features in that work occur at higher redshifts.
    Gaussian processes are defined such that a high degree of smoothness is imposed through the chosen kernel functions --- which typically result in an infinitely differentiable, and thus overly smooth, reconstructions \citep{2025arXiv250304273J}.
    In contrast, the free-form approach employed here does not impose such smoothness constraints, thereby allowing intricate, non-smooth structure in $w(a)$ to be more readily revealed.

    \subsection{DESI results}

    To determine which features were driven by which dataset, nested sampling runs were performed using each dataset individually.
    It was found that the DESI BAO are responsible for the higher-redshift feature in the joint reconstruction (see left panel of Figure~\ref{fig:desisdss}). 
    In this case, the deep BAO feature cannot admit a constant $w$; the evidence increases consistently up to $n=10$, although the overall flexknot (log-)evidence is still less than $\Lambda$CDM by $-0.71\pm0.18$.

    Much of the initial discussion following the DESI release \citep[e.g.,][]{lrgdiscussion1, lrgdiscussion2} focussed on the two LRG data points at $z=0.510$ and $0.706$, corresponding to scale factors of $a=0.662$ and $0.586$, respectively.
    Each point was removed in turn to assess whether one or the other was primarily responsible for the deviation from $\Lambda$CDM, and the pair were also replaced with the SDSS DR16 LRG data points.
    The reconstruction using both SDSS LRG points instead of the DESI ones is shown in the right four panels of Figure~\ref{fig:desisdss}.
    This data combination still favours $\Lambda$CDM over $w$CDM, and the corresponding value of $w_0=-1.03\pm0.14$ is very similar to the standard DESI results of $w_0=-1.01\pm0.14$.
    Furthermore, the overall flexknot evidence with the SDSS points is $-2.26\pm0.51$, significantly lower than the unaltered DESI data.
    This supports the interpretation that the DESI LRGs are driving the dynamical structure.
    Visually, the DESI $1\sigma$ contour does not admit a constant $w$, whereas the contour from the SDSS LRG data does.

    In Figure~\ref{fig:baodistances}, the functional posteriors of the cosmological distances are compared to those from the best-fit $\Lambda$CDM.
    The reconstructed shape tracks the seven DESI datapoints, with the LRG points positioned near the dramatic change from quintessance to phantom dark energy.
    When the LRG points are removed in turn, the lower-redshift LRG appears to be the primary driver for the quintessance phase.
    In the final column of Figure~\ref{fig:baodistances}, the two LRG points have been replaced by the equivalent SDSS DR16 points.
    This has a similar posterior to that obtained when excluding the lower-redshift LRG, implying that the more distant LRG is the point which disagrees with SDSS.
    In the absence of an understanding of the correlation between these measurements, the tension between the DESI and SDSS LRGs cannot be quantified.

    This DESI+SDSS combination should not be treated as a robust BAO dataset for further scientific use.
    Rather, we seek only to illustrate that the DESI LRGs exhibit characteristics of dynamical dark energy which the SDSS LRGs do not.

    \subsection{Supernova results}

    \begin{figure}
        \begin{center}
            \includegraphics[width=0.95\columnwidth]{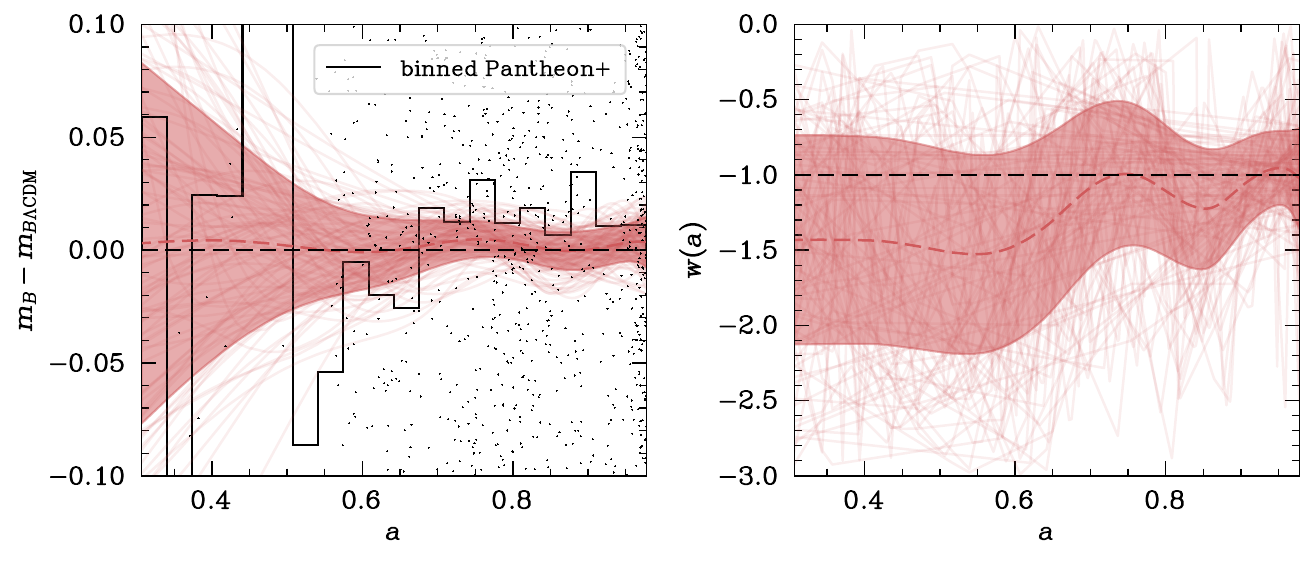}
        \end{center}
        \caption{
            Functional posterior of $m_B$ for Pantheon+ relative to $\Lambda$CDM.
            For clarity, the binned average is also shown because the variance of $m_B$ is much greater than the scale of the structure in the dark energy posterior.
        }\label{fig:pantheonmag}
    \end{figure}

    In the Pantheon+ reconstructions, a feature at lower redshifts is observed that is distinct from the DESI feature (see Figure~\ref{fig:pantheon}).
    Note that Pantheon+ shares a local maximum around $a\approx 0.75$ with DESI, an observation made clearer by Figure~\ref{fig:annaplot}.
    Figure~\ref{fig:pantheonmag} demonstrates how the reconstructed distance modulus $m_B$ deviates from $\Lambda$CDM.
    At redshifts greater than one (equivalently, at scale factors below one-half), the supernovae data become sparse, and the flexknot reconstruction reverts to the prior.
    Pantheon+ exhibits a preference for using as few knots as possible, leading to $\Lambda$CDM being favoured over flexknot dark energy.

    The shallower Pantheon+ feature almost permits a flat $w=-1$, resulting in flexknots being less favoured than $\Lambda$CDM, with a (log-)Bayes factor of $-2.81\pm0.63$.
    In contrast, the DES5Y posterior shows a much deeper phantom dip, which results in a slight preference for flexknots by $0.38\pm0.12$.
    This is shown in the right half of figure~\ref{fig:pantheon}.
    The $n=2$ model (CPL) is arguably unfairly penalised by the prior, as it is not possible for a single straight line to accommodate the structure in $w(a)$ while remaining above $w=-3$.

    It has been argued in \cite{georgedes5y} and \cite{ialowz} that low-redshift supernovae are primarily responsible for the evidence of dynamical dark energy.
    We defer a detailed investigation of the impact of low-redshift supernovae on the flexknot reconstruction of $w(a)$ to future work.

    \subsection{Tensions between BAO and supernovae}

    \begin{figure}
        \begin{center}
            \includegraphics[width=0.95\columnwidth]{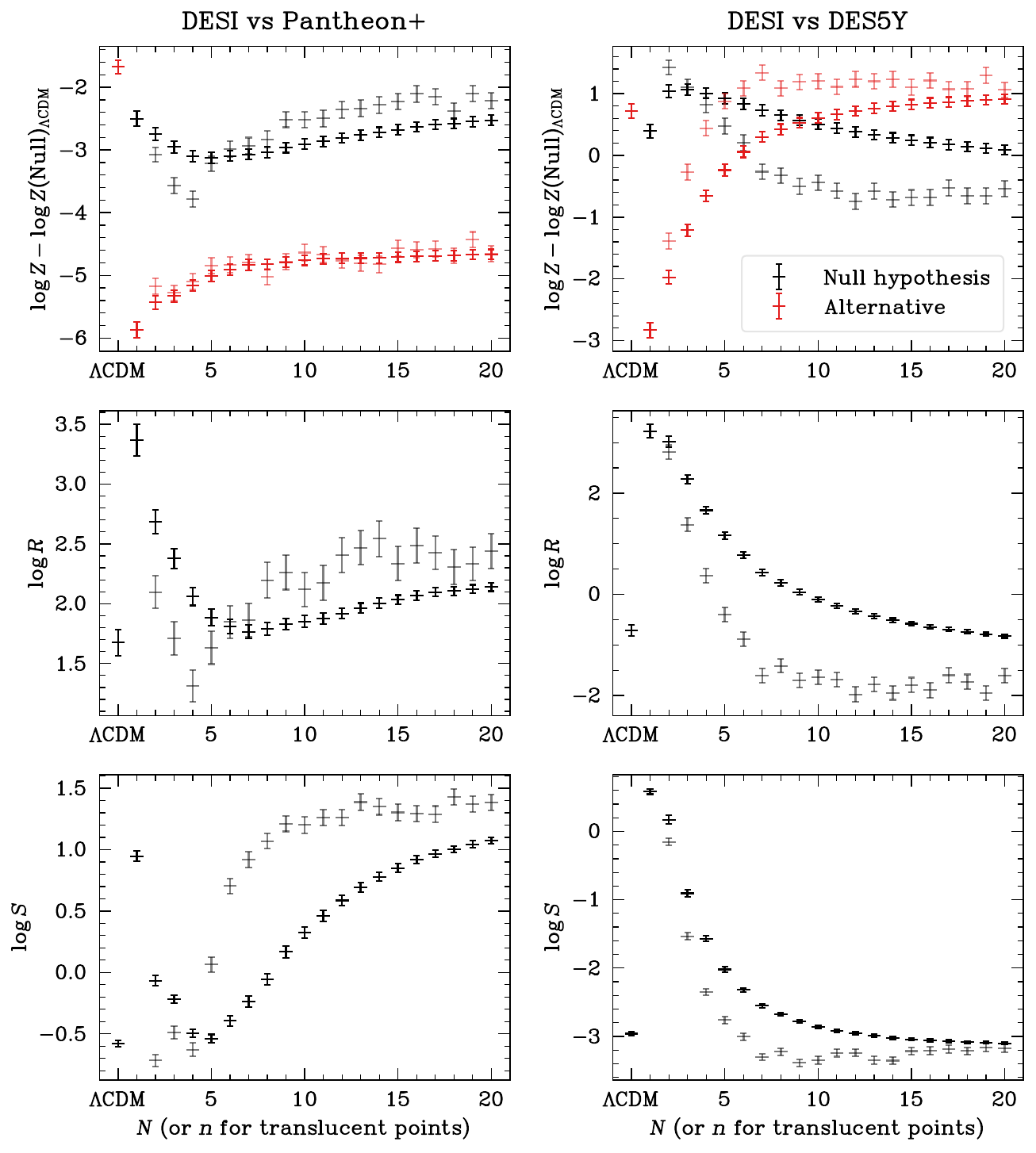}
        \end{center}
        \caption{
            Left column: evidences, Bayes ratios and suspiciousness values between DESI and Pantheon+ for both $\Lambda$CDM and flexknot dark energy.
            Translucent points correspond to specific $n$, and solid markers indicate cumulative results (combining all $n$ up to that value, excluding $\Lambda$CDM, i.e. $N$).
            Here, the null hypothesis is that the two datasets are measurements of the same universe, while the alternative hypothesis is that they are measurements of different cosmologies.
            The greatest tension (most negative $\log S$) is seen with $\Lambda$CDM.
            $w$CDM initially relives the tension, but moving to CPL increases it. 
            Figure~\ref{fig:desivspantheon2} shows that each dataset individually prefers a different straight-line fit.
            Increasing the number of knots beyond ten reduces the tension as the flexknot model better captures the structure present in both DESI and Pantheon+.
            Right column: The same analysis is shown for DESI and DES5Y supernovae.
            In this case, the tension statistics ($\log R$ and $\log S$) are substantially lower than for their Pantheon+ counterparts.
            Like Pantheon+, the model with the least tension is $n=1$, but $n=2$ is much closer.
            Increasing the number of knots only serves to increase disagreement, both suspiciousness and $R$-statistic values become increasingly negative.
        }\label{fig:tension}
    \end{figure}

    \begin{figure}
        \begin{center}
            \includegraphics[width=0.95\columnwidth]{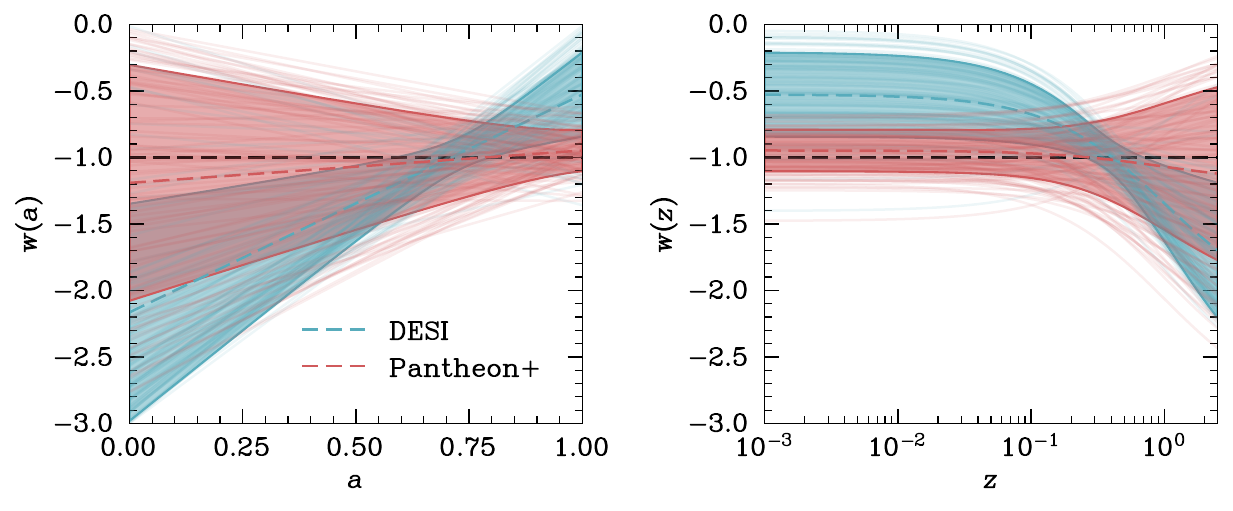}
        \end{center}
        \caption{
            Comparison of DESI and Pantheon+ for $n=2$, which corresponds to the CPL parameterisation (albeit with a different prior).
            It is apparent that each dataset favours different straight-line fits, which explains the dip in $\log R$ and $\log S$ in Figure~\ref{fig:tension}.
        }\label{fig:desivspantheon2}
    \end{figure}

    In Figure~\ref{fig:annaplot}, it was noted that the DESI and DES5Y reconstructions are incompatible at scale factors around $a\approx0.75$.
    To quantify this observation, the tension statistics introduced in Section~\ref{sec:tension} were computed for both cumulative and fixed numbers of knots.
    Figure~\ref{fig:tension} displays these statistics for DESI BAO compared with each of Pantheon+ and DES5Y.

    Between DESI and Pantheon+, $w$CDM improves their agreement compared to $\Lambda$CDM.
    However, the tension then worsens for $n=2$ (CPL), as Figure~\ref{fig:desivspantheon2} shows that each experiment individually favours different straight lines.
    Increasing the number of knots further subsequently reduces the tension again, as the flexknot is then able to capture the structure evident in each of the individual datasets.

    Replacing Pantheon+ with DES5Y supernovae tells a different story.
    $\log R$ and $\log S$ values are significantly lower than their Pantheon+ counterparts.
    In this case, $n=2$ results in the least tension, increasing the number of knots exacerbates the disagreement.
    This is consistent with the differences observed in the individual reconstructions.
    In Figure~\ref{fig:desisdss}, DESI prefers less negative $w$ at higher scale factors, while in Figure~\ref{fig:pantheon} DES5Y moves into a phantom regime at those same times.
    As flexknots are single-valued, no amount of additional flexibility can reconcile the two datasets, and adding more knots incurs an Occam penalty.

    \section{Conclusions and future work}
    \label{sec:conclusions}
    
    Bayesian evidence in favour of dynamical dark energy is not found for flexknot reconstructions of $w(a)$ relative to $\Lambda$CDM when combining DESI BAO with either Pantheon+ or DES5Y Type Ia supernovae.
    Only DES5Y taken in isolation shows a mild preference for dynamical dark energy.

    In both combinations of BAO and supernovae, a W-shaped structure is reconstructed in $w(a)$. The lower-redshift ``V'' is attributed to the supernovae, while the higher-redshift one is driven by the DESI BAO.
    At least in the case of DESI+Pantheon+, the interesting structure is not merely a symptom of unresolved disagreement between the datasets; this is supported by both the tension $R$-statistic and the suspiciousness values, which become increasingly positive as the number of knots increases.
    However, the DESI and DES5Y supernovae are incompatible at redshifts around $a\approx0.75$, which no amount of flexknot flexibility can reconcile.
    This is corroborated by the tension $R$-statistic and suspiciousness values that become increasingly negative with additional knots.

    However, if fewer knots had been considered, the conclusions may have been different. One, two, three, four and five knots are favoured by the DES5Y and DESI combination over $\Lambda$CDM, yielding positive $\log R$ values (albeit with negative suspiciousness).
    This observation highlights a flaw in marginalising tension statistics over multiple models, as the outcome is dependent on the model prior.

    While a direct tension analysis of the two supernova datasets is not possible, we conclude that the dynamical dark energy structure present in each supernova dataset does not agree with that in the other.

    It is also determined that the dynamical dark energy structure in the DESI BAO is largely caused by the two LRG points, particularly the one with effective redshift $z=0.510$.
    Similar results were obtained when that LRG is omitted or when both LRG-only points are replaced with their SDSS 2016 equivalents.
    The patched DESI+SDSS reconstruction clearly favours flexknot dark energy to a lesser extent than DESI DR1.

    In summary, no smoking gun for dynamical dark energy has been uncovered through the combined datasets.
    Each combination either lacks evidence for dynamical dark energy, or exhibits datasets incompatibility.
    However, it has been uncovered that flexknots with two, three or four knots are favoured over $\Lambda$CDM for the combination of DESI BAO plus DES5Y supernovae.
    Had a prior which favours fewer knots been chosen, the overall evidences for dynamical dark energy may also have been favoured.
    This suggests that structure present in $w(a)$ may be too complex to be captured with either $w$CDM or CPL dark energy.
    Certainly, it highlights a challenge of choosing an appropriate prior for statistics which include multiple models.
    
    The authors look forward to testing whether the upcoming second DESI data release will strengthen the structure in the $w(a)$ posteriors, or if the reign of $\Lambda$CDM will endure.

    \section*{Acknowledgements}

    This work was performed using the Cambridge Service for Data Driven Discovery (CSD3), part of which is operated by the University of Cambridge Research Computing on behalf of the STFC DiRAC HPC Facility (\url{www.dirac.ac.uk}).
    The DiRAC component of CSD3 was funded by BEIS capital funding via STFC capital grants ST/P002307/1 and ST/R002452/1 and STFC operations grant ST/R00689X/1.
    DiRAC is part of the National e-Infrastructure.

    The tension calculations in this work made use of \texttt{NumPy} \citep{numpy}, \texttt{SciPy} \citep{scipy}, and \texttt{pandas} \citep{pandaszenodo, pandaspaper}.
    The plots were produced in \texttt{matplotlib} \citep{matplotlib}, using the \texttt{smplotlib} template created by \citet{smplotlib}.

    The authors thank Toby Lovick and Suhail Dhawan for their assistance with the Ia supernovae likelihoods, and A.N.~Ormondroyd thanks A.L.~Jones for her constructive feedback on the plots.

    \section*{Data Availability}

    The pared-down Python pipeline and nested sampling chains used in this work can be obtained from Zenodo \citep{ormondroyd_2025_15025604}.



    \bibliographystyle{mnras}
    \bibliography{desi} 




    \appendix

    \section{Simplified pipeline}\label{apx:pipeline}
    \subsection{Cobaya versus custom pipeline}
    Due to their widespread use and ease of installation, the \texttt{Cobaya} pipeline is often employed for analyses like this work.
    \texttt{Cobaya} makes use of \texttt{CAMB} or \texttt{CLASS} to compute cosmological distances.
    Out of the box, no interface to \texttt{CAMB}'s \texttt{set\_w\_a\_table} method is provided by \texttt{Cobaya}, so this was added.
    The forced identifiability transform required for an efficient sorted prior for the $a_i$ was also implemented \citep{labelswitching}.

    In view of the surprising structure present in the $w(a)$ posteriors, an alternative pipeline stripped to the essentials was constructed as a consistency check.
    This has four advantages over using the \texttt{Cobaya} pipeline:
    \begin{itemize}
        \item Freedom to choose any prior.
        \item Confidence that the changes to \texttt{Cobaya} have been performed successfully.
        \item Reassurance is provided that the structure is not an artefact of data preparation for the \texttt{Cobaya} likelihoods (for example, the Pantheon+ redshift cut-off was only apparent upon inspection of the \texttt{Cobaya} source code).
        \item The simplicity of flexknots is leveraged so that a large part of the distance calculations can be carried out analytically.
    \end{itemize}

\begin{figure}
        \begin{center}
            \includegraphics[width=0.95\columnwidth]{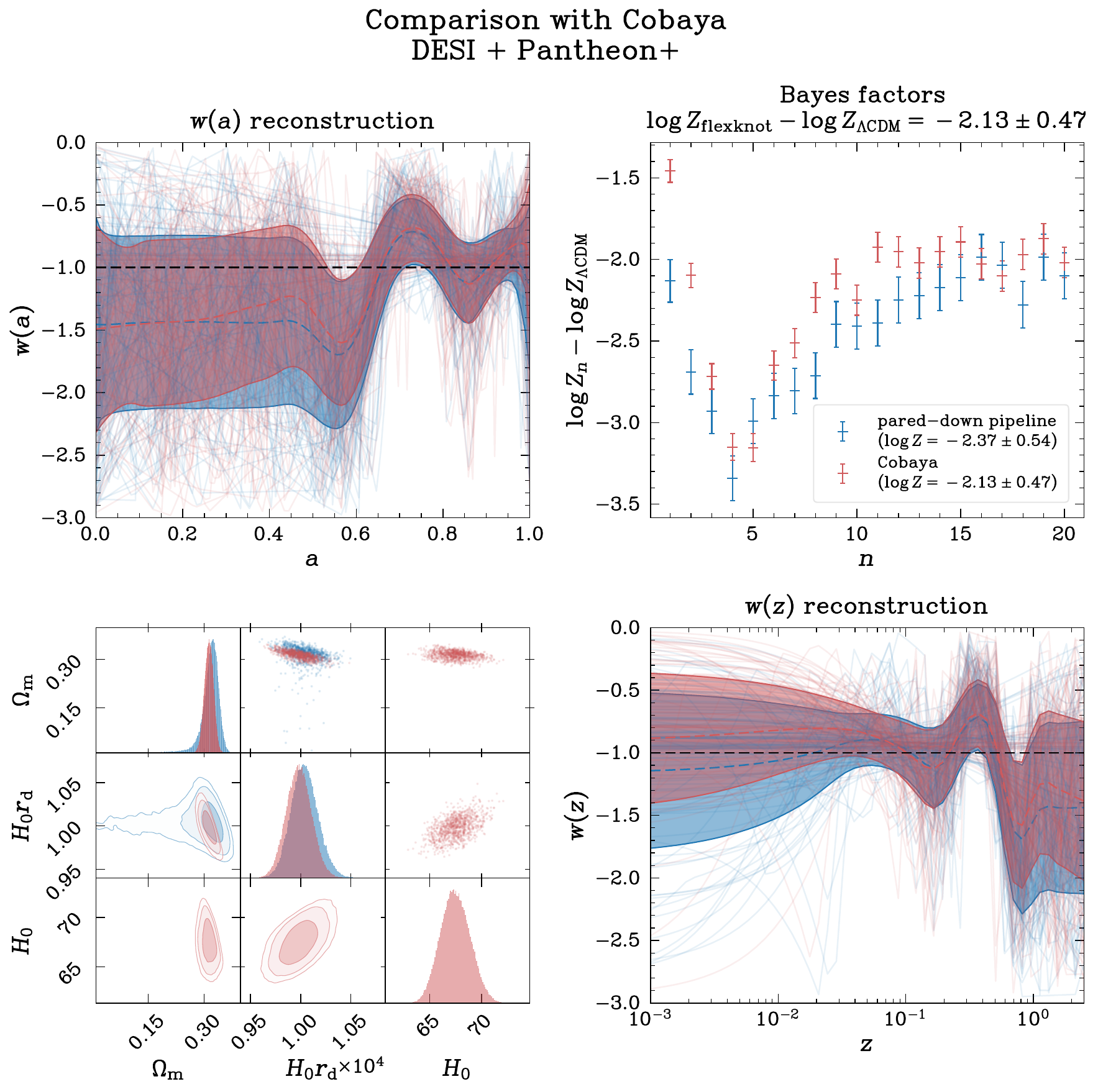}
        \end{center}
        \caption{
            A comparison is made between flexknot dark energy reconstructions using the \texttt{Cobaya} pipeline and the pipeline used in this work.
            In the upper-left panel, posterior samples of the evolution of $w(a)$ are shown, and the results from each pipeline are very similar.
            In the upper-right panel, the evidence for the $n^\text{th}$ flexknot is shown (normalised to the evidence of $\Lambda$CDM).
            The differences are expected because the \texttt{Cobaya} prior has four more parameters than the prior used in this work.
            In the lower-left panel, the posteriors for the parameters on which the likelihoods principally depend are shown.
            These are derived parameters in the \texttt{Cobaya} pipeline.
            In the lower-right panel, the same reconstructions transformed into $w(z)$ are shown.
            The pipeline used in this work employs a higher supernova redshift cut-off than \texttt{Cobaya}, but the $w$ posteriors are consistent.
            The \texttt{Cobaya} approach evidently contains $H_0$ information, as it is not present in either dataset.
            $H_0$ is not sampled in the pared-down pipeline, so it does not appear in those panels of the corner plot.
        }
        \label{fig:cobayadesipan}
    \end{figure}

    Since the computation of CMB power spectra is not required, all the distance calculations can be easily written in Python with little other than \texttt{NumPy} and \texttt{SciPy}.
    A comparison between results from the two pipelines is shown in Figure~\ref{fig:cobayadesipan}.

    Assuming zero curvature, the first Friedmann equation with an evolving $w$ is:
    \begin{equation}
        \begin{aligned}
            h^2(z) = \frac{H^2(z)}{H_0^2} &= \Omega_\text m(1+z)^3 + (1-\Omega_\text m) f_\text{DE}(z)\text,\\
            f_\text{DE}(z) &= \exp\left(3\int_0^z\frac{1+w(z')}{1+z'}\mathrm dz'\right)\text,\\
            w(z) &= \mathrm{flexknot}(1/(1+z), \theta_w)\text.
        \end{aligned}
    \end{equation}
    More detail on the integration of $f_\text{DE}$ is given in Section~\ref{apx:fde}.

\begin{figure}
        \begin{center}
            \includegraphics[width=0.95\columnwidth]{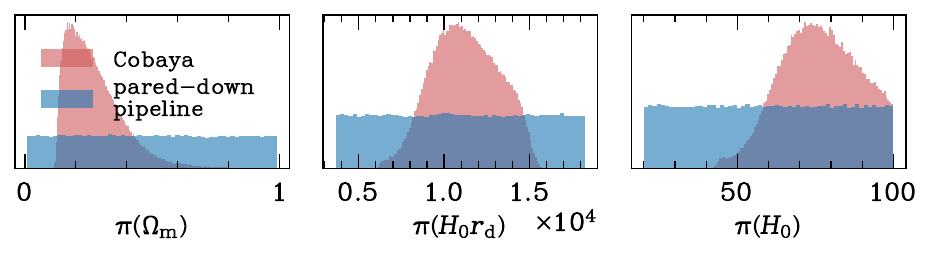}
        \end{center}
        \caption{
            Effective priors on $\Omega_\mathrm m$, $H_0r_\mathrm d$ and $H_0$ as used by the \texttt{Cobaya} pipeline are compared to the uniform priors used by the pipeline developed in this work for the combination of BAO and Type Ia supernovae.
        }\label{fig:priors}
    \end{figure}

    In the \texttt{Cobaya} pipeline, the user is forced to fix or sample the full set of six $\Lambda$CDM parameters to provide the necessary input to \texttt{CAMB}.
    By creating an independent pared-down pipeline for this work, the parameters of interest are sampled directly, with complete control to match exactly the priors used in \cite{desicrossingstatistics}.
    Figure~\ref{fig:priors} shows the difference between samples of the effective priors used in the \texttt{Cobaya} pipeline and the intended uniform priors. In practice, $H_0$ and $M_B$ are analytically marginalised and thus not sampled.

    Despite all the differences between the two pipelines, the character of the $w(a)$ reconstructions is found to be very similar between both pipelines, thereby supporting the claim that the structure is present in the data and not an artefact of the sampling pipeline.

    \subsection{Flexknot dark energy integrations}\label{apx:fde}

    For consistency with the notation of other dark energy parameterisations, $w_0$ is defined as the present-day value of $w$.
    Due to this choice, the labels of the flexknot parameters are assigned in reverse with respect to the scale factor. 

    With flexknot dark energy, each section of the dark energy flexknot is a straight-line segment in scale factor $a$, so $f_\mathrm{DE}$ can be computed analytically.
    Given the section between $a_{i+1}$ and $a_i$ has slope $m_i$ and intercept $c_i$, begin with the indefinite integral:
    \begin{equation}
        \begin{aligned}
            \int\frac{1+w(z)}{1+z}\mathrm dz &= \int\frac{1 + c_i + \frac{m_i}{1+z}}{1+z}\mathrm dz = \int \left(\frac{1 + c_i}{1+z} + \frac{m_i}{(1+z)^2}\right)\mathrm dz \\
            &= (1+c_i)\log(1+z) - m_i\frac{1}{1+z} + \text{const} \\
            &= -(1+c_i)\log a - m_i a + \text{const.}
        \end{aligned}
    \end{equation}
    Note that the definite integral from $0$ to $z$ corresponds to working backwards from $1$ down to $a$, the scale factor corresponding to $z$.
    The domain is split at the knots, which in redshift space occur at $z_i = 1/a_i - 1$, ending at the section containing $a$.
    \begin{equation}
        \begin{aligned}
            \int_0^z\frac{1+w(z')}{1+z'}\mathrm dz' &= \sum_{i=0}^{z_i<z} \big[-(1+c_i)\log a(z) - m_i a(z)\big]_{z=z_i}^{z_\text{upper}} \\
            &= \sum_{i=0}^{a_i > \frac{1}{1+z}} \big[(1+c_i)\log a + m_i a\big]^{a=a_i}_{a_\text{lower}}\text,\\
            a_\text{lower} &= \frac 1{1+z_\text{upper}} = \max\left(\frac{1}{1+z}, a_{i+1}\right),\\
            m_i &= \frac{w_i - w_{i+1}}{a_i - a_{i+1}}\text,\\
            c_i &= w_i - m_i a_i = \frac{w_{i+1}a_i - w_i a_{i+1}}{a_i - a_{i+1}}\text.
        \end{aligned}
    \end{equation}
    It should also be noted that the labels $i$ and $i+1$ may appear reversed, this is because the knots are numbered from the present day to the past.

    The function $h(z)$ can now be used to compute the cosmological distances needed for the BAO and supernovae likelihoods. 
    The integral $\int\frac{\mathrm dz}{h(z)}$ was computed numerically using \texttt{scipy.integrate.quad}, which employs a technique from the FORTRAN library \texttt{QUADPACK}.
    It was found that splitting the domain by the knots eliminated precision warnings, and improved speed fivefold.

    For Type Ia supernovae, it is inefficient to integrate to the hundreds of supernovae required for each $D_\text{L}(z)$ independently, so \texttt{quad} was used to integrate up to the closest supernova, then the remaining integrations were calculated from there using the trapezium rule.
    In contrast, only a handful of integrations are required for the BAO, so they were treated individually.

    \section{Marginalised supernova likelihoods}\label{apx:absolute}
    \subsection{Marginalisation over $M_B$}

    The supernova likelihood depends on the absolute magnitude $M_B$ of Type Ia supernovae, which is a nuisance parameter.
    The likelihood is given by
    \begin{equation}
        \begin{aligned}
            \mathcal L(D | \theta) &= \frac{1}{\sqrt{|2\pi\Sigma|}}\exp - \frac 1 2 \vec\Delta^T\Sigma^{-1}\vec\Delta\text,\\
            \vec\Delta &= (\vec m_B - M_B) - \mu(\vec z, \theta)\text,
        \end{aligned}
    \end{equation}
    where $\vec m_B$ denotes the observed magnitudes and $\vec z$ are the redshifts of the supernovae.
    $\mu$ is the calculated distance modulus at those redshifts given cosmology described by $\theta$.

    For brevity, define $\vec x = \vec m_B - \mu(\vec z, \theta)$, and let $M_B$ be itself multiplied by a vector of ones, so that
    \begin{equation}
        \begin{aligned}
            M_B &= \vec M = M\vec 1 = M \begin{bmatrix} 1 \\ 1 \\ \vdots\end{bmatrix} \text,\\
            \mathcal L(\vec m_B, \vec z, M_B|\theta) &= \frac{1}{\sqrt{|2\pi\Sigma|}}\exp\left(-\frac{1}{2}(\vec x - M\vec 1)^T\Sigma^{-1}(\vec x - M\vec 1)\right) \text.\\
        \end{aligned}
    \end{equation}

    Assuming the prior on $M_B$ is uniform with volume $V_{M_B} = {M_B}_\text{max} - {M_B}_\text{min}$ and wide enough that the posterior tends to zero at the edges, the likelihood is analytically marginalised over $M_B$:
    \begin{equation}
        \begin{aligned}
            &\mathcal L(\vec m_B, \vec z|\theta) = \int_{{M_B}_\text{min}}^{{M_B}_\text{max}} \frac 1 V_{M_B} \mathcal L(\vec m_B, \vec z, M_B|\theta)\mathrm dM_B \\
            &=\frac{1}{V_{M_B}\sqrt{|2\pi\Sigma|}}\int_{-\infty}^{\infty}\exp\left(-\frac{1}{2}(\vec x - M\vec 1)^T\Sigma^{-1}(\vec x - M\vec 1)\right)\mathrm dM \\
            &= \frac{1}{V_{M_B}\sqrt{|2\pi\Sigma|}}e^{-\frac{1}{2}\vec x^T\Sigma^{-1}\vec x}\\
            &\times\int_{-\infty}^{\infty}e^{
                -\frac 1 2 \left(\vec 1^T\Sigma^{-1}\vec 1\right)M^2
                +\frac 1 2 (\vec 1^T\Sigma^{-1}\vec x + \vec x^T\Sigma^{-1}\vec 1) M
                }\mathrm dM \text.\\
        \end{aligned}
    \end{equation}
    By denoting the coefficients of $M^2$ and $M$ in the integral $a$ and $b$ respectively, the integral can be solved by completing the square:
    \begin{equation}
        \begin{aligned}
            &\mathcal L(\vec m_B, \vec z|\theta) = \\
            &=\frac{1}{V_{M_B}\sqrt{|2\pi\Sigma|}}e^{-\frac{1}{2}\vec x^T\Sigma^{-1}\vec x}
            \int_{-\infty}^{\infty}\exp\left[- \frac 1 2 (a M^2 - bM) \right]\mathrm dM \\
            &= \frac{1}{V_{M_B}\sqrt{|2\pi\Sigma|}}e^{-\frac{1}{2}\vec x^T\Sigma^{-1}\vec x} \int_{-\infty}^{\infty}\exp\left[-\frac 1 2 a\left[\left(M - \frac {b}{2a}\right)^2 - \frac{b^2}{4a^2}\right]\right]\mathrm dM \\
            &= \frac{1}{V_{M_B}\sqrt{|2\pi\Sigma|}}e^{-\frac{1}{2}\vec x^T\Sigma^{-1}\vec x + \frac{b^2}{8a}}\sqrt{\frac{2\pi}{a}} \text.
        \end{aligned}
    \end{equation}

    The two terms in $b$ are scalar and mutual transposes, thus they are equal and can be interchanged freely.
    This permits the vector $\vec x$ to be moved to the outside:
    \begin{equation}
        \begin{aligned}
            &\mathcal L(\vec m_B, \vec z|\theta) = \\
            &= \frac 1 V_{M_B} \sqrt{\frac{2\pi}{|2\pi\Sigma|a}}\exp - \frac 1 2 \left[\vec x^T\Sigma^{-1}\vec x - \frac 1 a \left(\frac{\vec 1^T\Sigma^{-1}\vec x + \vec x^T\Sigma^{-1}\vec 1}{2}\right)^2\right] \\
            &= \frac 1 V_{M_B} \sqrt{\frac{2\pi}{|2\pi\Sigma|a}} \exp - \frac 1 2 \left[\vec x^T\Sigma^{-1}\vec x - \frac 1 a (\vec x^T\Sigma^{-1}\vec 1)(\vec 1^T\Sigma^{-1}\vec x)\right] \\
            &= \frac 1 V_{M_B} \sqrt{\frac{2\pi}{|2\pi\Sigma|\vec 1^T\Sigma^{-1}\vec 1}} \exp - \frac 1 2 \vec x^T\tilde\Sigma^{-1}\vec x \text,\\
            \tilde\Sigma^{-1} &= \Sigma^{-1} - \frac{\Sigma^{-1}\vec 1\vec 1^T\Sigma^{-1}}{\vec 1^T\Sigma^{-1}\vec 1} \text.
            \label{eq:marginalisedmb}
        \end{aligned}
    \end{equation}
    On the final line, the definition $a = \vec 1^T\Sigma^{-1}\vec 1$ was reinserted.

    \subsection{Marginalisation over $H_0$}
    \label{apx:h0}
    $H_0$ may also be marginalised from supernova likelihoods.
    A uniform prior is assumed for $H_0$:
    \begin{equation}
        \pi(H_0) = \begin{cases}
            \frac{1}{H_{0\text{max}} - H_{0\text{min}}} = \frac{1}{V_{H_0}} & \text{if } H_{0\text{min}}\leq H_0\leq H_{0\text{max}} \text,\\
            0 & \text{otherwise}\text.
        \end{cases}
    \end{equation}
    The vector $\vec x$ from the previous subsection is separated into $\vec x = h - \vec y$, where the dependence on redshift and magnitude data is contained in $\vec y$. In particular,
    \begin{equation}
        \begin{aligned}
            x &= \vec m_B - \mu(\vec z, \theta) = \vec m_B - 5\log_{10}{\frac{D_\mathrm L(\vec z)}{\SI{10}{pc}}}\\
            &= \vec m_B - 5\log_{10}{(1+\vec z_\mathrm{hel})\int_0^{\vec z_\mathrm{HD}}\frac{\mathrm dz'}{h(z')}} - 5\log_{10}{\left(\frac{c}{\SI{10}{pc}H_0}\right)}\\
            &= 5\log_{10}\left(\frac{\SI{10}{pc}H_0}{c}\right) - \vec y = h - \vec y\text.
        \end{aligned}
    \end{equation}
    Focus attention on the exponent in equation \ref{eq:marginalisedmb}:
    \begin{equation}
        \begin{aligned}
            & -\frac 1 2 (\vec y-h)^T\tilde\Sigma^{-1}(\vec y-h) \\
            = &-\frac 1 2 \vec y^T\tilde\Sigma^{-1}\vec y + \frac 1 2 \vec y^T\tilde\Sigma^{-1}\vec 1 + \frac 1 2 h \vec 1^T\tilde\Sigma^{-1}\vec y - \frac 1 2 h \vec 1^T\tilde\Sigma^{-1}\vec 1 h \\
            = &-\frac 1 2 \vec y^T\tilde\Sigma^{-1}\vec y + \vec y^T\tilde\Sigma^{-1}\vec 1 h\text,
        \end{aligned}
    \end{equation}
    where it is used that $\tilde\Sigma^{-1}$ is symmetric so that $\vec 1^T\tilde\Sigma^{-1}\vec y = \vec y^T\tilde\Sigma^{-1}\vec 1$, and it is noted $\vec 1^T\tilde\Sigma^{-1}\vec 1$ identically vanishes:
    \begin{equation}
        \vec 1^T\tilde\Sigma^{-1}\vec 1 = \vec 1^T\Sigma^{-1}\vec 1 - \frac{\vec 1^T\Sigma^{-1}\vec 1\vec 1^T\Sigma^{-1}\vec 1}{\vec 1^T\Sigma^{-1}\vec 1} = 0\text.
    \end{equation}
    
    Focussing on that second term, logarithms are converted to base $e$, the exponential returns and integration over $H_0$ is carried out:

    \begin{equation}
        \begin{aligned}
            & \int_{H_{0\text{min}}}^{H_{0\text{max}}} \frac{\mathrm dH_0}{V_{H_0}}\exp\left[\vec y^T\tilde\Sigma^{-1}\vec 1\times5\log_{10}\left(\frac{\SI{10}{pc}H_0}{c}\right)\right]\\
            & = \int_{u_\text{min}}^{u_\text{max}}\frac{\mathrm du}{V_u} e^{Y\log u} = \int_{u_\text{min}}^{u_\text{max}} \frac{\mathrm du}{V_u} u^Y\\
            & = \frac{1}{V_u}\frac{(u_\text{max})^{Y+1}-(u_\text{min})^{Y+1}}{Y+1}
            \text,\\
            &Y = \vec y^T\tilde\Sigma^{-1}\vec 1\times \frac{5}{\log 10}\text,\quad u = \frac{\SI{10}{pc}H_0}{c}\text.
        \end{aligned}
    \end{equation}
    The pieces may now be put all together to compute the marginalised supernova likelihood.
    The final result is
    \begin{equation}
        \begin{aligned}
            \mathcal L(\vec m_B, \vec z) &= \frac{c}{\SI{10}{pc}V_{H_0}V_{M_B}}\sqrt{\frac{2\pi}{|2\pi\Sigma|\vec 1^T\tilde\Sigma^{-1}\vec 1}}\exp -\frac 1 2 \vec y^T\tilde\Sigma^{-1}\vec y \\
            &\times \frac{(u_\text{max})^{Y+1}-(u_\text{min})^{Y+1}}{Y+1}
        \end{aligned}
    \end{equation}

    \section{Supernova redshift cut-off}\label{apx:iacutoff}
\begin{figure}
        \begin{center}
            \includegraphics[width=0.95\columnwidth]{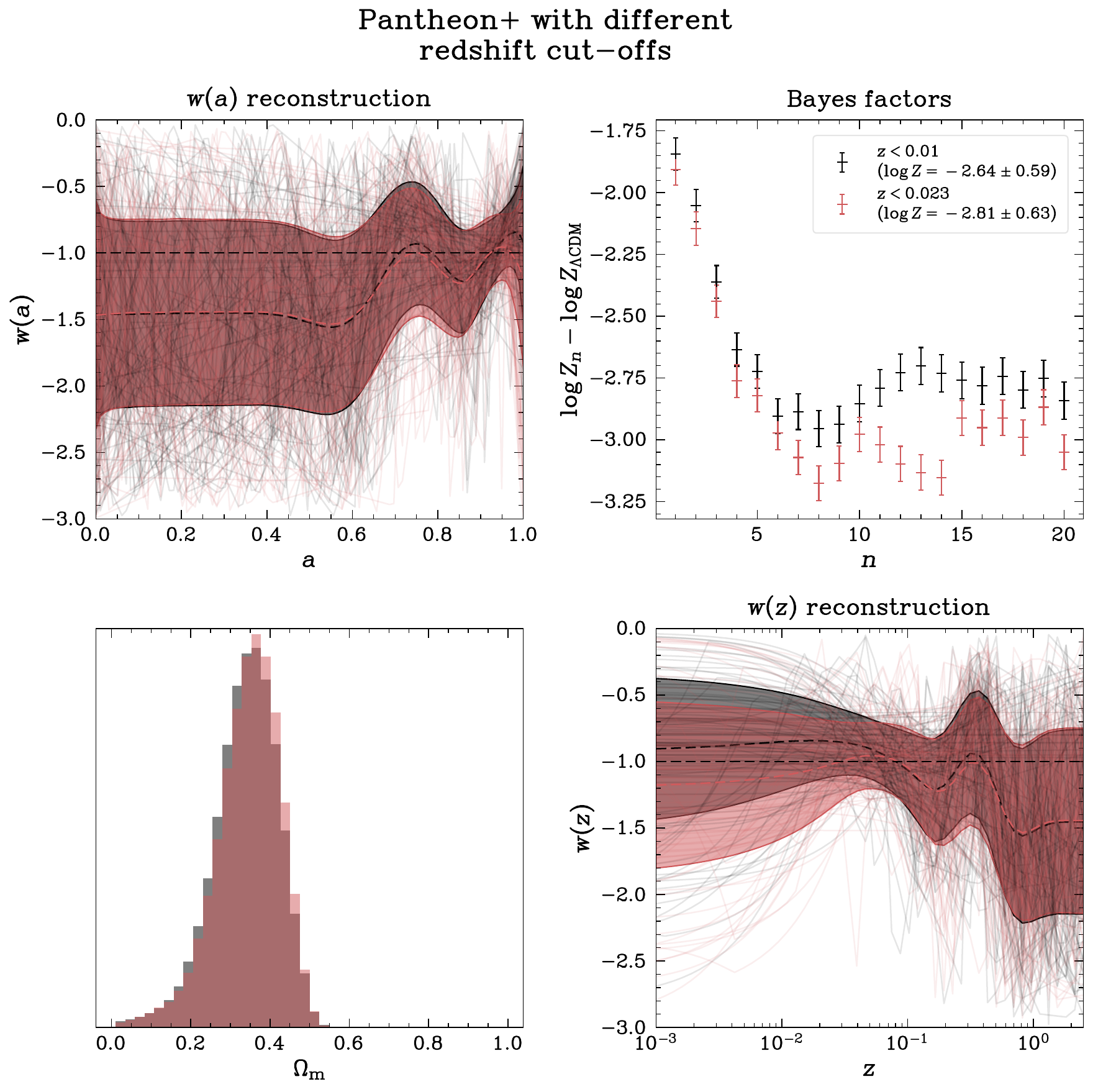}
        \end{center}
        \caption{
            Comparison of Pantheon+ supernovae with lower redshift cut-offs of $z=0.01$ and $z=0.023$.
            Their character is identical.
        }\label{fig:iacutoff}
    \end{figure}

    Different analyses of the Pantheon+ dataset have used different redshift cut-offs for the supernovae. Figure~\ref{fig:iacutoff} shows that the character of the reconstructions is identical for the two popular cut-offs of $z=0.01$ and $z=0.023$.

    \section{Tension statistics over multiple models}\label{apx:tension}

    The target expression is an equivalent for equation \ref{eq:suspiciousness} (repeated here for clarity):
    \begin{equation}
        \log S = \langle \log \mathcal L \rangle_{\mathcal P(A, B)} - \langle \log \mathcal L \rangle_{\mathcal P(A)} - \langle \log \mathcal L \rangle_{\mathcal P(B)} \text.
    \end{equation}

    \subsection{Suspiciousness for multiple models}
    A series of models, indexed by $i$, is considered, each characterised by its parameters $\theta_i$ and its own separate nested sampling run with a corresponding evidence.
    For flexknots, $i$ corresponds to the total number of knots, which is fixed at values from $1$ to $N$ in turn, but this formalism can be applied to any set of models.
    To compute the tension between two datasets averaged over these models, the statistics from the individual nested sampling runs are combined as follows.

    Using $\theta_i$ to refer to the parameters of the model with $i$ knots, the total evidence for the model with up to $N$ knots is the individual evidences $Z_i$ weighted by the model selection prior $\pi(n=i)$:
    \begin{equation}
        \begin{aligned}
            Z &= \int \mathcal L(\theta)\pi(\theta)\mathrm d\theta = \sum_{i=1}^N \int \mathcal L(\theta_i)\pi(\theta_i|n=i)\pi(n=i)\mathrm d\theta_i \\
            &= \sum_{i=1}^N \pi(n=i)Z_i \text.
        \end{aligned}
    \end{equation}
    Conditional probability has been used to separate $n$ from the other parameters by writing $\pi(\theta) = \pi(n=i)\pi(\theta_i | n=i)$, and the likelihood does not directly depend on $n$, so $\mathcal L(\theta) = \mathcal L(\theta_i)$.

    In order to compute suspiciousness, it is efficient to consider $\langle \log \mathcal L \rangle_\mathcal P$ directly rather than compute $\log Z + \mathcal D_\mathrm{KL}(\mathcal P||\pi)$.
    With a uniform prior $\pi(n=i) = 1/N$, the model selection posterior is weighted by evidence:
    \begin{equation}
        \mathcal P(n=i) = \frac{\pi(n=i) Z_i} {\sum_j \pi(n=j) Z_j} = \frac{Z_i} {\sum_j Z_j} \text,\\
    \end{equation}
    which can be used to compute the average log-likelihood over all numbers of knots:
    \begin{equation}
        \begin{aligned}
            \langle \log \mathcal L \rangle_{\mathcal P} &= \int \mathcal P \log \mathcal L \mathrm d\theta \\
            &= \sum_{i=1}^N \int \mathcal P(\theta_i|n=i)\mathcal P(n=i) \log \mathcal L \mathrm d\theta_i \\
            &= \frac{\sum_i Z_i \int \mathcal P (\theta_i|n=i)\log\mathcal L \mathrm d\theta_i} {\sum_j Z_j} = \frac{\sum_i Z_i \langle\log\mathcal L\rangle_{\mathcal P_i}} {\sum_j Z_j}\text.
        \end{aligned}
    \end{equation}

    It is confirmed in Appendix~\ref{apx:klconsistency} that the above formulae for evidence and $\langle\log\mathcal L\rangle_\mathcal P$ are consistent with the similarly modified KL divergence.
    Armed with these formulae, tension suspiciousness between two datasets marginalised over several different models can be computed.

    \subsection{KL divergence consistency check}
    \label{apx:klconsistency}

    For completeness and to verify that the formulae for calculating the evidence and $\langle\log\mathcal L\rangle_\mathcal P$ from those of the individual flexknot nested sampling runs are consistent, the KL divergence is computed directly:
    \begin{equation}
        \begin{aligned}
            &\mathcal D_\mathrm{KL}(\mathcal P || \pi) = \int \mathcal P(\theta) \log\frac{\mathcal P(\theta)}{\pi(\theta)} \mathrm d\theta \\
            &= \sum_i \int \mathcal P(\theta_i|n=i)\mathcal P(n=i) \log\frac{\mathcal P(\theta_i|n=i)\mathcal P(n=i)}{\pi(\theta_i|n=i)\pi(n=i)} \mathrm d\theta_i \text.\\
        \end{aligned}
    \end{equation} 
    The prior on the flexknot parameters is conditioned on $n$, so that $\pi(\theta) = \pi(n=i)\pi(\theta_i|n=i)$, and the model-selection posterior is given by $\mathcal P(n=i) = \frac{\pi(n=i)Z_i}{\sum_j \pi(n=j)Z_j}$.
    Denote $\pi_i = \pi(n=i)$ for brevity. Then,
    \begin{equation}
        \begin{aligned}
            &= \sum_i \int \mathcal P(\theta_i|n=i)\frac{\pi_iZ_i}{\sum_j \pi_jZ_j} \log\left(\frac{\mathcal P(\theta_i|n=i)}{\pi(\theta_i|n=i)}\frac{\frac{\pi_iZ_i}{\sum_j\pi_jZ_j}}{\pi_i}\right)\mathrm d\theta_i \\
            &\begin{aligned}
                = \frac{1}{\sum_j\pi_jZ_j} \sum_i \pi_iZ_i &\left[\int\mathcal P(\theta_i|n=i)\log\frac{\mathcal P(\theta_i|n=i)}{\pi(\theta_i|n=i)}\mathrm d\theta_i\right.\\
                +&\left.\log\frac{Z_i}{\sum_j \pi_jZ_j}\int\mathcal P(\theta_i|n=i)\mathrm d\theta_i\right]\text.\\
            \end{aligned}
        \end{aligned}
    \end{equation}
    It is recalled from Bayes' theorem that $\mathcal P(\theta_i|n=i) / \pi(\theta_i|n=i) = \mathcal L(\theta_i) / Z_i$, and since the posterior $\mathcal P(\theta_i|n=i)$ is normalised over $\theta_i$, it follows that
    \begin{equation}
        \begin{aligned}
            = \frac{1}{\sum_j\pi_jZ_j} \sum_i \pi_iZ_i
            \left[\int \mathcal P(\theta_i|n=i)\log\frac{\mathcal L(\theta_i)}{Z_i}\mathrm d\theta_i \right.\\
            + \left.\log\frac{Z_i}{\sum_j \pi_jZ_j}\right]\text.\\
            = \frac{1}{\sum_j\pi_jZ_j} \sum_i \pi_iZ_i\left[\langle\log\mathcal L\rangle_{\mathcal P_i}- \int \mathcal P(\theta_i|n=i)\log Z_i\mathrm d\theta_i\right.\\
            + \left.\log Z_i - \log\sum_j \pi_jZ_j\right]\text.
        \end{aligned}
    \end{equation}
    The middle two terms in the brackets cancel as $Z_i$ does not depend on $\theta_i$ and the posterior once again is normalised.
    The final term does not depend on $i$ so it can be factored out of the sum, which then cancels with the denominator:
    \begin{equation}
        = \frac{\sum\pi_iZ_i\langle\log\mathcal L\rangle_{\mathcal P_i}}{\sum\pi_jZ_j} - \log\sum_i\pi_iZ_i \equiv \langle\log\mathcal L\rangle_{\mathcal P} - \log Z\text.
    \end{equation}
    Finally, the two terms in this expression are identified with those calculated in Appendix~\ref{apx:tension}, thereby confirming the consistency of the formulae.

    \section{Irrelevant parameters}\label{apx:anthony}

    There may be sampled parameters on which the model, and thus the likelihood, does not actually depend.
    Consider a model with parameters $\theta$, evidence $Z = \int\mathcal L(\theta)\pi(\theta)\mathrm d\theta$, where the data $D$ have been suppressed for brevity.
    An extra parameter $\varphi$ is introduced, on which neither the model nor the likelihood depend, the evidence remains unchanged:

    \begin{equation}
        \begin{aligned}
            &\mathcal L(\theta, \varphi) = \mathcal L(\theta), \quad \pi(\theta, \varphi) = \pi(\theta)\pi(\varphi)\\
            &\int \mathcal L(\theta, \varphi)\pi(\theta, \varphi)\mathrm d\theta \mathrm d\varphi = \int L(\theta)\pi(\theta)\mathrm d\theta \int\pi(\varphi)\mathrm d\varphi = Z \text,
        \end{aligned}
    \end{equation}
    where the fact the prior on $\theta$ is normalised is used.


    \bsp	
    \label{lastpage}
\end{document}